\documentclass[a4paper, 11pt, oneside]{article}
\usepackage[a4paper,inner=2cm,outer=2cm,top=3cm,bottom=2.5cm]{geometry}

\usepackage[tbtags]{amsmath}
\usepackage{amssymb}
\usepackage{amsthm}
\usepackage{dsfont}
\usepackage{slashed}
\usepackage{mathrsfs}
\usepackage[tbtags]{mathtools}
\usepackage{color}

\usepackage[export]{adjustbox}

\usepackage[english]{babel}

\usepackage[utf8x]{inputenc}

\usepackage{graphicx,epsfig}

\usepackage{setspace}

\usepackage{booktabs}

\usepackage{float}

\usepackage{dcolumn}

\usepackage{nextpage} 

\usepackage{newclude}

\usepackage[small,bf]{caption}
\usepackage{subcaption}

\usepackage{url}

\usepackage[colorlinks=true,
	linkcolor=blue,
	citecolor=blue,
	urlcolor=blue,
	filecolor=black
]{hyperref}

\usepackage[T1]{fontenc}

\usepackage{cancel}

\usepackage{tabularx}

\usepackage{pifont}

\usepackage{etex}

\usepackage{appendix}

\usepackage[numbers,sort&compress]{natbib}

\usepackage{placeins}

\usepackage{ragged2e}

\usepackage{ifmtarg}

\usepackage{tikz}
\usetikzlibrary{shapes,arrows}

\usepackage{bchart}

\usepackage{gensymb}

\renewcommand{\Re}{\text{Re}\,}

\newcommand{\<}{\langle}
\renewcommand{\>}{\rangle}

\newcommand{\BR}{\text{BR}}

\newcommand{\beq}{\begin{equation}}
\newcommand{\eeq}{\end{equation}}
\newcommand{\GeV}{\,\text{GeV}}
\newcommand{\MeV}{\,\text{MeV}}
\newcommand{\keV}{\,\text{keV}}

\newcommand{\Fpi}{F_\pi}
\newcommand{\M}{\mathcal{M}}
\newcommand{\F}{\mathcal{F}}
\newcommand{\Q}{\mathcal{Q}}
\newcommand{\diff}{d}

\newcommand{\eps}{\epsilon}
\newcommand{\Tr}{\text{Tr}}
\newcommand{\Order}{\mathcal{O}}

\newcommand{\remark}[1]{}

\makeatletter
\newcommand{\Cr}[2]{\@ifmtarg{#2}{\mathcal{C}_{#1}}{\mathcal{C}_{#1}\big[#2\big]}}
\makeatother

\newcommand{\nn}{\nonumber\\}

\newcolumntype{.}{D{.}{.}{2} } 
\newcolumntype{d}{D{.}{.}{2.2} }
\newcolumntype{L}[1]{>{\RaggedRight\hspace{0pt}}p{#1}}
\newcolumntype{R}[1]{>{\RaggedLeft\hspace{0pt}}p{#1}}

\makeatletter
  \def\my@tag@font{\normalsize}
  \def\maketag@@@#1{\hbox{\m@th\normalfont\my@tag@font#1}}
  \let\amsmath@eqref\eqref
  \renewcommand\eqref[1]{{\let\my@tag@font\relax\amsmath@eqref{#1}}}
\makeatother

\makeatletter
\renewcommand\paragraph{\@startsection{paragraph}{4}{\z@}%
  {-3.25ex \@plus -1ex \@minus -0.2ex}%
  {0.01pt}%
  {\bfseries}%
}
\makeatother

\makeatletter
\def\@xfootnote[#1]{%
  \protected@xdef\@thefnmark{#1}%
  \@footnotemark\@footnotetext}
\makeatother

\usepackage{abstract}

\allowdisplaybreaks[1]

\begin{document}

\mbox{}

\vspace{-1.75cm}
\hfill{}\begin{minipage}[t][0cm][t]{5cm}
\raggedleft
\footnotesize
INT-PUB-20-015
\end{minipage}
\vspace{1.25cm}

\bigskip

\begin{center}
{\LARGE{\bf Asymptotic behavior of meson transition form factors}}

\vspace{0.5cm}

Martin Hoferichter${}^{a,b}$, Peter Stoffer${}^{c}$

\vspace{1em}

\begin{center}
\it
\mbox{} \\
${}^a$Albert Einstein Center for Fundamental Physics, Institute for Theoretical Physics, \\ University of Bern, Sidlerstrasse 5, CH--3012 Bern, Switzerland \\
\mbox{}\\
${}^b$Institute for Nuclear Theory, University of Washington, Seattle, WA 98195-1550, USA \\
\mbox{}\\
${}^c$Department of Physics, University of California at San Diego, \\ 9500 Gilman Drive, La Jolla, CA 92093-0319, USA
\end{center} 

\end{center}

\vspace{3em}

\hrule

\begin{abstract}
One of the open issues in evaluations of the contribution from hadronic light-by-light scattering to the anomalous magnetic moment of the muon $(g-2)_\mu$ concerns the role of heavier scalar, axial-vector, and tensor-meson intermediate states. The coupling of axial vectors to virtual photons is suppressed for small virtualities by the Landau--Yang theorem, but otherwise there are few rigorous constraints on the corresponding form factors. In this paper, we first derive the Lorentz decomposition of the two-photon matrix elements into scalar functions following the general recipe by Bardeen, Tung, and Tarrach. Based on this decomposition, we then calculate the asymptotic behavior of the meson transition form factors from a light-cone expansion in analogy to the asymptotic limits for the pseudoscalar transition form factor derived by Brodsky and Lepage. Finally, we compare our results to existing data as well as previous models employed in the literature.
\end{abstract}

\hrule


\setcounter{tocdepth}{3}
\tableofcontents

\numberwithin{equation}{section}


\section{Introduction}

The asymptotic behavior of pseudoscalar transition form factors (TFFs)---describing the decay of a pseudoscalar meson into two (virtual) photons $P\to\gamma^*(q_1)\gamma^*(q_2)$---has been studied in detail in the literature using an expansion along the light cone $x^2=0$, with the central result that at leading order the corresponding TFF, e.g.\ for the pion, can be expressed as~\cite{Lepage:1979zb,Lepage:1980fj,Brodsky:1981rp}
\beq
\label{eq:pQCD}
F_{\pi^0\gamma^*\gamma^*}(q_1^2,q_2^2)=-\frac{2\Fpi}{3}\int_0^1\diff u\frac{\phi_\pi(u)}{u q_1^2+(1-u) q_2^2} +\Order\big(q_i^{-4}\big),
\eeq
in terms of the decay constant $\Fpi=92.28(19)\MeV$~\cite{Tanabashi:2018oca} and the wave function $\phi_\pi(u)$.
This approach has been widely applied both for kinematic configurations that follow from a strict operator product expansion (OPE), in particular, the symmetric limit~\cite{Nesterenko:1982dn,Novikov:1983jt}
\beq
\label{eq:OPE_limit}
F_{\pi^0\gamma^*\gamma^*}(q^2,q^2)=-\frac{2\Fpi}{3q^2}+\Order\big(q^{-4}\big),
\eeq
but also for the singly-virtual case
\beq
\label{eq:BL_limit}
F_{\pi^0\gamma^*\gamma^*}(q^2,0)=-\frac{2\Fpi}{q^2}+\Order\big(q^{-4}\big).
\eeq
The latter is obtained by formal evaluation of~\eqref{eq:pQCD} for the asymptotic form of the wave function $\phi_\pi(u)=6u(1-u)$ and is often referred to as 
the Brodsky--Lepage (BL) limit of the singly-virtual TFF. As pointed out in~\cite{Gorsky:1987,Manohar:1990hu}, this goes beyond a strict OPE, in the sense that the wave-function approach already resums higher-order terms. Moreover, considerable effort has been devoted to extending the leading-order result~\eqref{eq:pQCD}
including $\alpha_s$ corrections~\cite{delAguila:1981nk,Braaten:1982yp} and higher-order terms in the context of QCD sum rules~\cite{Chernyak:1981zz,Chernyak:1983ej,Radyushkin:1996tb,Khodjamirian:1997tk,Agaev:2010aq,Stefanis:2012yw,Agaev:2014wna,Mikhailov:2016klg}.
Results by the BaBar experiment for the singly-virtual pion TFF for space-like virtualities above $10\GeV^2$~\cite{Aubert:2009mc} suggested  
that there could be substantial corrections to the BL limit, while later data by the Belle collaboration~\cite{Uehara:2012ag} did not point to a similarly fast rise of the TFF.
Moreover, the BaBar measurement of the $\eta$, $\eta'$ TFFs~\cite{BABAR:2011ad} proved in better agreement with the BL expectation, although in this case the detailed interpretation depends on singlet corrections and $\eta$--$\eta'$ mixing patterns. 

In recent years, these constraints on the asymptotic behavior of pseudoscalar TFFs have become vital ingredients for determinations of the contribution from pseudoscalar intermediate states to hadronic light-by-light scattering (HLbL) in the anomalous magnetic moment of the muon $(g-2)_\mu$. In fact, with the contribution of various hadronic intermediate states organized in terms of dispersion relations~\cite{Hoferichter:2013ama,Colangelo:2014dfa,Colangelo:2014pva,Colangelo:2015ama,Colangelo:2017qdm,Colangelo:2017fiz}, 
the pseudoscalar poles are completely determined by the respective TFFs. For the pion, the TFF has, in turn, been reconstructed from dispersion relations~\cite{Niecknig:2012sj,Schneider:2012ez,Hoferichter:2012pm,Hoferichter:2014vra,Hoferichter:2017ftn,Hoferichter:2018dmo,Hoferichter:2018kwz}, leading to a result for the pion-pole contribution in perfect agreement with calculations using Canterbury approximants~\cite{Masjuan:2017tvw}, lattice QCD~\cite{Gerardin:2019vio}, and Dyson--Schwinger equations~\cite{Eichmann:2019tjk,Raya:2019dnh}, and a similar program exists for the $\eta$, $\eta'$ poles~\cite{Stollenwerk:2011zz,Hanhart:2013vba,Kubis:2015sga,Xiao:2015uva,Kubis:2018bej}. In either case, asymptotic constraints on the TFF are critical in controlling the high-energy part of the $g-2$ integral. This aspect becomes particularly important when matching to short-distance constraints~\cite{Melnikov:2003xd,Bijnens:2019ghy,Colangelo:2019uex,Colangelo:2019lpu}. 

Going beyond pseudoscalar poles, the second most important intermediate states are $2\pi$~\cite{Colangelo:2017qdm,Colangelo:2017fiz}, which require input on the amplitudes for $\gamma^*\gamma^*\to\pi\pi$~\cite{GarciaMartin:2010cw,Hoferichter:2011wk,Moussallam:2013una,Danilkin:2018qfn,Hoferichter:2019nlq,Danilkin:2019opj}. However, some resonances in the $2\pi$ system, such as the $f_0(980)$ or the $f_2(1270)$, should be reasonably well described by a narrow-width approximation (NWA), in which case information on the respective TFFs would again be required. Moreover, for higher-multiplicity intermediate states, such as $3\pi$, a NWA may be the only realistic way to estimate their contribution, given the complexity of the dispersion relations for a general three-particle intermediate state. Again, the TFFs would be key input quantities. Phenomenologically, there is some information on the TFFs of scalar ($f_0(980)$~\cite{Mori:2007bu,Uehara:2008ep,Uehara:2009cka,Masuda:2015yoh}, $a_0(980)$~\cite{Uehara:2009cf}, $f_0'(1370)$~\cite{Uehara:2010mq}, $a_0(1450)$~\cite{Uehara:2009cf}), axial-vector ($f_1(1285)$~\cite{Gidal:1987bn,Aihara:1988uh,Achard:2001uu}, $f_1'(1420)$~\cite{Achard:2007hm}), and tensor ($f_2(1270)$~\cite{Mori:2007bu,Uehara:2008ep,Uehara:2009cka,Masuda:2015yoh}, $a_2(1320)$~\cite{Behrend:1989rc,Albrecht:1996mr,Acciarri:1997rb}, $f_2'(1525)$~\cite{Albrecht:1989re,Acciarri:2000ex,Abe:2003vn,Uehara:2013mbo}, $a_2'(1700)$~\cite{Acciarri:2000ex,Abe:2003vn}) mesons, but in neither case is the data situation comparable to the pseudoscalar TFFs. For the axial-vector resonances, an additional complication arises due to the Landau--Yang theorem~\cite{Landau:1948kw,Yang:1950rg}, which forbids a coupling to two on-shell photons. Finally, constraints on these TFFs can be obtained when assuming the saturation of $\gamma\gamma$ sum rules with narrow resonances~\cite{Pascalutsa:2012pr,Danilkin:2016hnh}. Existing estimates for the contribution to HLbL scattering from such heavy intermediate states rely on the available phenomenological information~\cite{Pauk:2014rta} and/or further input from the matching to short-distance constraints~\cite{Melnikov:2003xd,Jegerlehner:2017gek,Knecht:2018sci}, resonance chiral theory~\cite{Roig:2019reh}, 
or holographic models~\cite{Leutgeb:2019gbz,Cappiello:2019hwh}. 

In all cases, however, the resulting estimates are still quite model dependent, with major issues including---apart from the obvious scarcity of data---ambiguities in the definition of resonance contributions (so far always taken from a Lagrangian formulation), 
kinematic singularities in the TFF decomposition, and assumptions on the asymptotic behavior. In this paper we will address the latter two. First, for use in HLbL scattering a TFF decomposition is required that is free of kinematic singularities, which does not apply to the standard decomposition~\cite{Poppe:1986dq,Schuler:1997yw} formulated in terms of helicity components (in the case of scalar and tensor mesons, the decompositions in~\cite{Kopp:1973hp} are free of kinematic singularities, but no proof is provided). In Sect.~\ref{sec:BTT} we will therefore derive Lorentz decompositions, following the general recipe established by Bardeen, Tung, and Tarrach (BTT)~\cite{Bardeen:1969aw,Tarrach:1975tu}, that are manifestly free of kinematic singularities. Second, the only available constraints on the asymptotic behavior of the resulting TFFs originate from the quark model of~\cite{Schuler:1997yw}, but even there only for a subset of the TFFs, as well as for one particular limit of the  scalar and axial-vector TFFs from the OPE~\cite{Knecht:2018sci,Roig:2019reh} and from holographic QCD~\cite{Leutgeb:2019gbz}. Scalar~\cite{Cheng:2005nb,Lu:2006fr,Kroll:2016mbt}, axial-vector~\cite{Yang:2005gk,Yang:2007zt}, and tensor~\cite{Braun:2000cs,Cheng:2010hn,Braun:2016tsk} mesons have been studied using light-cone techniques, with some early work already in~\cite{Kopp:1973hp}. However, we are not aware of representations analogous to~\eqref{eq:pQCD}, certainly not in a basis useful for HLbL scattering. We will fill this gap in Sect.~\ref{sec:BL}. Some phenomenological applications will be discussed in Sect.~\ref{sec:data}, before closing with a summary and outlook in Sect.~\ref{sec:summary}.


\section{Lorentz structure and helicity amplitudes}
\label{sec:BTT}

\subsection[Pseudoscalar mesons: $J^{PC} = 0^{-+}$]{\boldmath Pseudoscalar mesons: $J^{PC} = 0^{-+}$}

To define notation and conventions, 
we first consider the well-known case of a pseudoscalar meson decaying into two off-shell photons. The meson $P$ is treated as an asymptotic state, i.e., in the NWA we have:
\begin{align}
	\begin{split}
		\< \gamma^*(q_1,\lambda_1) & \gamma^*(q_2,\lambda_2) | P(p) \> \\
			&= - e^2 {\epsilon_\mu^{\lambda_1}}^*(q_1) {\epsilon_\nu^{\lambda_2}}^*(q_2) \int d^4x \, d^4y \, e^{i (q_1 \cdot x + q_2 \cdot y )} \< 0 | T \{ j_\mathrm{em}^\mu(x) j_\mathrm{em}^\nu(y) \} | P(p) \> \\
			&= - e^2 {\epsilon_\mu^{\lambda_1}}^*(q_1) {\epsilon_\nu^{\lambda_2}}^*(q_2) \int d^4x \, d^4y \, e^{i q_1 \cdot x} e^{i (q_1 + q_2 - p) \cdot y} \< 0 | T \{ j_\mathrm{em}^\mu(x) j_\mathrm{em}^\nu(0) \} | P(p) \> \\
			&= - (2\pi)^4 \delta^{(4)}(q_1+q_2-p) e^2 {\epsilon_\mu^{\lambda_1}}^*(q_1) {\epsilon_\nu^{\lambda_2}}^*(q_2) \int d^4x \, e^{i q_1 \cdot x} \< 0 | T \{ j_\mathrm{em}^\mu(x) j_\mathrm{em}^\nu(0) \} | P(p) \> \\
			&= i (2\pi)^4 \delta^{(4)}(q_1+q_2-p) e^2 {\epsilon_\mu^{\lambda_1}}^*(q_1) {\epsilon_\nu^{\lambda_2}}^*(q_2) \M^{\mu\nu}(p \rightarrow q_1,q_2),
	\end{split}
\end{align}
where we have introduced the $T$-matrix elements
\begin{align}
	\M^{\mu\nu}(p \rightarrow q_1,q_2) = i \int d^4x \, e^{i q_1 \cdot x} \< 0 | T \{ j_\mathrm{em}^\mu(x) j_\mathrm{em}^\nu(0) \} | P(p) \> 
\end{align}
involving the electromagnetic current
\beq
j_\mathrm{em}^\mu(x)=\bar q(x)\Q \gamma^\mu q(x),\qquad q=(u,d,s)^T,\qquad \Q=\frac{1}{3}\text{diag}(2,-1,-1).
\eeq
The helicity amplitudes are defined by
\begin{align}
	H_{\lambda_1\lambda_2} = {\epsilon_\mu^{\lambda_1}}^*(q_1) {\epsilon_\nu^{\lambda_2}}^*(q_2) \M^{\mu\nu}(q_1,q_2) .
\end{align}
We define polarization vectors in the rest frame of the meson as
\begin{align}
		\label{eq:PolarizationVectors}
		\epsilon_\pm(q_1) &= \mp \frac{1}{\sqrt{2}} ( 0, 1, \pm i, 0 ) ,  &
		\epsilon_0(q_1) &= \frac{1}{\xi_1} ( |\vec q| , 0, 0, E_1 ) , \notag\\
		\epsilon_\pm(q_2) &= \mp \frac{1}{\sqrt{2}} ( 0, 1, \mp i, 0 ) , &
		\epsilon_0(q_2) &= \frac{1}{\xi_2} ( - |\vec q| , 0, 0, E_2 ) .
\end{align}
The momenta satisfy $p=q_1+q_2$. In the meson rest frame, they are given by
\beq
		q_1 = (E_1, 0, 0, |\vec q| ) , \qquad
		q_2 = (E_2, 0, 0, -|\vec q| ) , \qquad
		p = (m_P, 0, 0, 0 ) ,
\eeq
where
\beq
		E_1 = \sqrt{ q_1^2 + |\vec q|^2} = \frac{m_P^2 + q_1^2 - q_2^2}{2m_P}, \qquad E_2 = \sqrt{ q_2^2 + |\vec q|^2} = \frac{m_P^2 - q_1^2 + q_2^2}{2m_P} ,\qquad 
		|\vec q| = \frac{\lambda_{P12}^{1/2}}{2 m_P},
\eeq
and the K\"all\'en function is defined by $\lambda_{P12} := \lambda(m_P^2, q_1^2, q_2^2)$, $\lambda(a,b,c) = a^2 + b^2 + c^2 - 2(ab + bc + ca)$.

In the pseudoscalar case, finding the decomposition of $\M^{\mu\nu}$ into scalar amplitudes that are free of kinematic singularities is trivial, since there is only a single gauge-invariant Lorentz structure that can be constructed, leading to the conventional parameterization in terms of the pseudoscalar TFF $F_{P\gamma^*\gamma^*}(q_1^2,q_2^2)$ according to
\beq
\label{TFF_pion}
\M_{\mu\nu}=\epsilon_{\mu\nu\alpha\beta}q_1^{\alpha}q_2^{\beta} F_{P\gamma^*\gamma^*}(q_1^2,q_2^2),
\eeq
where $\epsilon^{0123}=+1$.
Its normalization is related to the
on-shell decay width $\Gamma_{\gamma\gamma}$ by
\beq
|F_{P\gamma^*\gamma^*}(0,0)\big|^2=\frac{4}{\pi\alpha^2m_P^3}\Gamma_{\gamma\gamma}.
\eeq

\subsection[Scalar mesons: $J^{PC} = 0^{++}$]{\boldmath Scalar mesons: $J^{PC} = 0^{++}$}

For scalar mesons the Lorentz decomposition of the matrix element $\M^{\mu\nu}$ becomes slightly less straightforward because now there are two independent structures that need to be chosen in such a way that both are free of kinematic singularities. 
To illustrate the general procedure in the more complicated axial-vector and tensor cases, we apply already here the BTT recipe. First, crossing symmetry requires
\begin{align}
	\M^{\mu\nu}(q_1,q_2) = \M^{\nu\mu}(q_2,q_1)
\end{align}
and parity conservation forbids the presence of an epsilon tensor. The elementary building blocks are therefore $g^{\mu\nu}$, $q_1^\mu$, $q_2^\mu$, i.e.
\begin{align}
	\{ L_i^{\mu\nu} \} = \{  g^{\mu\nu} , q_1^\mu q_1^\nu , q_1^\mu q_2^\nu , q_2^\mu q_1^\nu , q_2^\mu q_2^\nu \} .
\end{align}
Next, we impose gauge invariance by contracting these structures with the projector
\begin{align}
	I^{\mu\nu} = g^{\mu\nu} - \frac{q_2^\mu q_1^\nu}{q_1 \cdot q_2} ,
\end{align}
which satisfies
\begin{align}
	q_1^\mu I_{\mu\nu} = 0 , \qquad q_2^\nu I_{\mu\nu} = 0 , \qquad I_{\mu\mu'} \M^{\mu'\nu} = {\M_\mu}^{\nu} , \qquad I_{\nu'\nu} \M^{\mu\nu'} = {{\M^\mu}_\nu} .
\end{align}
Hence, we calculate the contracted Lorentz structures
\begin{align}
	\bar L_i^{\mu\nu} = I^{\mu\mu'} I^{\nu'\nu} {L_{i,\mu'\nu'}} .
\end{align}
Three structures project to zero. We then remove the kinematic singularities by taking linear combinations and multiplying the irreducible poles by $q_1 \cdot q_2$. This leads to the following gauge-invariant structures:
\begin{align}
	\begin{split}
		T_1^{\mu\nu} &= q_1 \cdot q_2 g^{\mu\nu} - q_2^\mu q_1^\nu , \\
		T_2^{\mu\nu} &= q_1^2 q_2^2 g^{\mu\nu} + q_1 \cdot q_2 q_1^\mu q_2^\nu - q_1^2 q_2^\mu q_2^\nu - q_2^2 q_1^\mu q_1^\nu .
	\end{split}
\end{align}
We define the photon crossing operator as
\begin{align}
	\Cr{12}{f} = f(\mu \leftrightarrow \nu, q_1 \leftrightarrow q_2) .
\end{align}
The Lorentz structures are both symmetric under crossing:
\begin{align}
	\Cr{12}{T_{1,2}^{\mu\nu}} = T_{1,2}^{\mu\nu} .
\end{align}
Finally, to obtain dimensionless form factors $\F_i^S$, we define the Lorentz decomposition of the amplitude as:
\begin{align}
	\M^{\mu\nu} = \frac{1}{m_S} T_1^{\mu\nu} \F_1^S + \frac{1}{m_S^3} T_2^{\mu\nu} \F_2^S .
\end{align}

Further, contracting the Lorentz structures with the polarization vectors and evaluating the expression in the rest frame of the meson, the only non-vanishing helicity amplitudes, fulfilling $\lambda_1 = \lambda_2$, become
\begin{align}
	\begin{split}
		H_{++} = H_{--} &= -\frac{m_S^2-q_1^2-q_2^2}{2m_S} \F_1^S - \frac{q_1^2 q_2^2}{m_S^3} \F_2^S , \\
		H_{00} &= \frac{q_1^2 q_2^2}{\xi_1\xi_2} \left( - \frac{1}{m_S} \F_1^S - \frac{m_S^2 - q_1^2 - q_2^2}{2m_S^3} \F_2^S \right) .
	\end{split}
\end{align}
The differential decay width for the process $S(p) \to \gamma^*(q_1,\lambda_1) \gamma^*(q_2,\lambda_2)$ is given by
\begin{align}
	d\Gamma_{\gamma^*\gamma^*} = \frac{e^4}{32\pi^2} | H_{\lambda_1\lambda_2} |^2 \frac{\lambda_{S12}^{1/2}}{2m_S^3} d\Omega .
\end{align}
In terms of the form factors, we obtain for the decay width summed over $\lambda_{1,2}$ (with $\xi_1 = \sqrt{q_1^2}$, $\xi_2 = \sqrt{q_2^2}$)
\begin{align}
	\begin{split}
		\Gamma_{\gamma^*\gamma^*} = \frac{e^4}{16\pi} \frac{\lambda_{S12}^{1/2}}{m_S} &\bigg( \frac{\lambda_{S12}+6q_1^2q_2^2}{2m_S^4} |\F_1^S|^2 + \frac{q_1^2 q_2^2(\lambda_{S12}+12q_1^2q_2^2)}{4m_S^8} |\F_2^S|^2 \\
			& + \frac{3q_1^2q_2^2(m_S^2-q_1^2-q_2^2)}{m_S^6} \Re\big( \F_1^S {\F_2^S}^* \big)\bigg) .
	\end{split}
\end{align}
Therefore, the normalization of $\F_1^S$ is given by the on-shell width (a factor of $1/2$ in $\Gamma_{\gamma\gamma}$ with respect to $\Gamma_{\gamma^*\gamma^*}$ is introduced for indistinguishable on-shell photons):
\begin{align}
	| \F_1^S(0,0) |^2 &= \frac{4}{\pi \alpha^2 m_S} \Gamma_{\gamma\gamma} .
\end{align}

\subsection[Axial-vector mesons: $J^{PC} = 1^{++}$]{\boldmath Axial-vector mesons: $J^{PC} = 1^{++}$}
\label{sec:BTTAxials}

In close analogy to the (pseudo-) scalar case we define for the axial-vector mesons 
\begin{align}
	\begin{split}
		\< \gamma^*(q_1,\lambda_1) & \gamma^*(q_2,\lambda_2) | A(p,\lambda_A) \> \\
			&= - (2\pi)^4 \delta^{(4)}(q_1+q_2-p) e^2 {\epsilon_\mu^{\lambda_1}}^*(q_1) {\epsilon_\nu^{\lambda_2}}^*(q_2) \int d^4x \, e^{i q_1 \cdot x} \< 0 | T \{ j_\mathrm{em}^\mu(x) j_\mathrm{em}^\nu(0) \} | A(p,\lambda_A) \> \\
			&= i (2\pi)^4 \delta^{(4)}(q_1+q_2-p) e^2 {\epsilon_\mu^{\lambda_1}}^*(q_1) {\epsilon_\nu^{\lambda_2}}^*(q_2) \M^{\mu\nu}(\{p,\lambda_A\} \rightarrow q_1,q_2) \\
			&= i (2\pi)^4 \delta^{(4)}(q_1+q_2-p) e^2 {\epsilon_\mu^{\lambda_1}}^*(q_1) {\epsilon_\nu^{\lambda_2}}^*(q_2) \epsilon_\alpha^{\lambda_A}(p) \M^{\mu\nu\alpha}(q_1,q_2),
	\end{split}
\end{align}
with $T$-matrix elements
\begin{align}
	\M^{\mu\nu}(\{p,\lambda_A\} \rightarrow q_1,q_2) = \epsilon_\alpha^{\lambda_A}(p) \M^{\mu\nu\alpha}(q_1,q_2) = i \int d^4x \, e^{i q_1 \cdot x} \< 0 | T \{ j_\mathrm{em}^\mu(x) j_\mathrm{em}^\nu(0) \} | A(p,\lambda_A) \> .
\end{align}
The helicity amplitudes are defined by
\begin{align}
	H_{\lambda_1\lambda_2;\lambda_A} = {\epsilon_\mu^{\lambda_1}}^*(q_1) {\epsilon_\nu^{\lambda_2}}^*(q_2) \epsilon_\alpha^{\lambda_A}(p) \M^{\mu\nu\alpha}(q_1,q_2),
\end{align}
with photon polarization vectors as given in~\eqref{eq:PolarizationVectors}. We define the polarization vectors of the axial-vector meson as
\beq
		\label{eq:PolarizationVectorsAxial}
		\epsilon_\pm(p) = \mp \frac{1}{\sqrt{2}} ( 0, 1, \pm i, 0 ) , \qquad
		\epsilon_0(p) = ( 0 , 0, 0, 1 ).
\eeq

For the BTT decomposition of $\M^{\mu\nu\alpha}$ we first note that crossing symmetry requires
\begin{align}
	\M^{\mu\nu\alpha}(q_1,q_2) = \M^{\nu\mu\alpha}(q_2,q_1)
\end{align}
and that due to parity all structures need to involve one epsilon tensor. We write
\begin{align}
	\M^{\mu\nu\alpha} = \epsilon_{\beta\gamma\delta\eta} T^{\mu\nu\alpha\beta\gamma\delta\eta}
\end{align}
and construct the tensor $T^{\mu\nu\alpha\beta\gamma\delta\eta}$ with the elementary building blocks $g^{\mu\nu}$, $q_1^\mu$, $q_2^\mu$. A priori, the structures
\begin{align}
	g g g q , \quad g g q q q , \quad g q q q q q , \quad q q q q q q q,
\end{align}
have to be considered, but due to the antisymmetry of the epsilon tensor the last two structures immediately contract to zero. From the first two structures, we find the following possibilities:
\begin{align}
	\begin{split}
		\{ L_i^{\mu\nu\alpha} \} = \epsilon_{\beta\gamma\delta\eta} \{ &  q_1^\beta g^{\mu\gamma} g^{\nu\delta} g^{\alpha\eta} , q_2^\beta g^{\mu\gamma} g^{\nu\delta} g^{\alpha\eta} , 
													 q_1^\mu g^{\nu\beta} g^{\alpha\gamma} q_1^\delta q_2^\eta ,  q_2^\mu g^{\nu\beta} g^{\alpha\gamma} q_1^\delta q_2^\eta , \\
													& q_1^\nu g^{\mu\beta} g^{\alpha\gamma} q_1^\delta q_2^\eta ,  q_2^\nu g^{\mu\beta} g^{\alpha\gamma} q_1^\delta q_2^\eta , 
													 q_1^\alpha g^{\nu\beta} g^{\mu\gamma} q_1^\delta q_2^\eta ,  q_2^\alpha g^{\nu\beta} g^{\mu\gamma} q_1^\delta q_2^\eta  \} ,
	\end{split}
\end{align}
hence the set of naive Lorentz structures consists of eight elements
\begin{align}
	\begin{split}
		\{ L_i^{\mu\nu\alpha} \} = \{ &   \epsilon^{\mu\nu\alpha\beta} {q_1}_\beta ,  \epsilon^{\mu\nu\alpha\beta} {q_2}_\beta , 
													 \epsilon^{\nu\alpha\beta\gamma} q_1^\mu {q_1}_\beta {q_2}_\gamma , \epsilon^{\nu\alpha\beta\gamma} q_2^\mu {q_1}_\beta {q_2}_\gamma , \\
													& \epsilon^{\mu\alpha\beta\gamma} q_1^\nu {q_1}_\beta {q_2}_\gamma , \epsilon^{\mu\alpha\beta\gamma} q_2^\nu {q_1}_\beta {q_2}_\gamma , 
													 \epsilon^{\mu\nu\beta\gamma} q_1^\alpha {q_1}_\beta {q_2}_\gamma , \epsilon^{\mu\nu\beta\gamma} q_2^\alpha {q_1}_\beta {q_2}_\gamma   \} .
	\end{split}
\end{align}
Next, we impose gauge invariance by contracting these structures with the gauge projector $I^{\mu\nu}$. Two structures project to zero. We then remove the kinematic singularities by taking linear combinations and multiplying the irreducible poles by $q_1 \cdot q_2$. This leads to the following set of structures:
\begin{align}
	\begin{split}
		\{ \bar T_i^{\mu\nu\alpha} \} = \Big\{
					& \epsilon^{\alpha \mu \beta\gamma} {q_1}_\beta {q_2}_\gamma q_1^\nu + \epsilon^{\alpha\mu\nu\beta} {q_1}_\beta q_1 \cdot q_2 , \\
					& \epsilon^{\alpha\nu\beta\gamma} {q_1}_\beta {q_2}_\gamma q_2^\mu + \epsilon^{\alpha\mu\nu\beta} {q_2}_\beta q_1 \cdot q_2 , \\
					& \epsilon^{\mu\nu\beta\gamma} {q_1}_\beta {q_2}_\gamma ( q_1^\alpha + q_2^\alpha ) , \\
					& \epsilon^{\mu\nu\beta\gamma} {q_1}_\beta {q_2}_\gamma ( q_1^\alpha - q_2^\alpha ) , \\
					& \epsilon^{\alpha\nu\beta\gamma} {q_1}_\beta {q_2}_\gamma q_1^\mu + \epsilon^{\alpha\mu\nu\beta} {q_2}_\beta q_1^2 , \\
					& \epsilon^{\alpha\mu\beta\gamma} {q_1}_\beta {q_2}_\gamma q_2^\nu + \epsilon^{\alpha\mu\nu\beta} {q_1}_\beta q_2^2
					\Big\} .
	\end{split}
\end{align}
In fact, these structures are not linearly independent due to the Schouten identity. We find the linear relations
\begin{align}
	\begin{split}
		\bar T_1^{\mu\nu\alpha} &= - \frac{1}{2} \bar T_3^{\mu\nu\alpha} - \frac{1}{2} \bar T_4^{\mu\nu\alpha} + \bar T_5^{\mu\nu\alpha} , \\
		\bar T_2^{\mu\nu\alpha} &= \frac{1}{2} \bar T_3^{\mu\nu\alpha} - \frac{1}{2} \bar T_4^{\mu\nu\alpha} + \bar T_6^{\mu\nu\alpha} .
	\end{split}
\end{align}
Finally, in any observable, the tensor will be contracted with
\begin{align}
	\label{eq:AxialPolarizationSum}
	s^A_{\alpha\alpha'}(p) := \sum_{\lambda_A} \epsilon_\alpha^{\lambda_A}(p) \epsilon_{\alpha'}^{\lambda_A}(p)^* = - \left( g_{\alpha\alpha'} - \frac{p_\alpha p_{\alpha'}}{m_A^2} \right) ,
\end{align}
which projects $\bar T_3^{\mu\nu\alpha}$ to zero. Hence, the third structure does not contribute to physical quantities and can be dropped. Therefore, we arrive at the final set of gauge-invariant Lorentz structures:
\begin{align}
	\begin{split}
		\{ T_i^{\mu\nu\alpha} \} = \Big\{
					& \epsilon^{\mu\nu\beta\gamma} {q_1}_\beta {q_2}_\gamma ( q_1^\alpha - q_2^\alpha ) , \\
					& \epsilon^{\alpha\nu\beta\gamma} {q_1}_\beta {q_2}_\gamma q_1^\mu + \epsilon^{\alpha\mu\nu\beta} {q_2}_\beta q_1^2 , \\
					& \epsilon^{\alpha\mu\beta\gamma} {q_1}_\beta {q_2}_\gamma q_2^\nu + \epsilon^{\alpha\mu\nu\beta} {q_1}_\beta q_2^2
					\Big\} .
	\end{split}
\end{align}
The Lorentz structures transform under photon crossing as
\begin{align}
\label{crossing_axial}
	\Cr{12}{T_1^{\mu\nu\alpha}} = - T_1^{\mu\nu\alpha} , \quad \Cr{12}{T_2^{\mu\nu\alpha}} = - T_3^{\mu\nu\alpha} .
\end{align}
We define dimensionless form factors $\F_i^A$, which are the scalar functions in the Lorentz decomposition of the amplitude:
\begin{align}
	\label{eq:AxialFormFactors}
	\M^{\mu\nu\alpha} = \frac{i}{m_A^2} \sum_{i=1}^3 T_i^{\mu\nu\alpha} \F_i^A(q_1^2, q_2^2) .
\end{align}

In terms of these form factors, the helicity amplitudes become
\begin{align}
	\begin{split}
		H_{++;0} = - H_{--;0} &= \frac{\lambda_{A12}}{2m_A^3} \F_1^A - \frac{q_1^2(m_A^2-q_1^2+q_2^2)}{2m_A^3} \F_2^A - \frac{q_2^2(m_A^2+q_1^2-q_2^2)}{2m_A^3} \F_3^A , \\
		H_{+0;+} = - H_{-0;-} &= \frac{q_1^2 q_2^2}{\xi_2 m_A^2} \F_2^A + \frac{q_2^2(m_A^2-q_1^2-q_2^2)}{2\xi_2 m_A^2} \F_3^A , \\
		H_{0+;-} = - H_{0-;+} &= - \frac{q_1^2(m_A^2-q_1^2-q_2^2)}{2\xi_1 m_A^2} \F_2^A - \frac{q_1^2 q_2^2}{\xi_1 m_A^2} \F_3^A ,
	\end{split}
	\label{hel_axial}
\end{align}
where $\lambda_{A12} := \lambda(m_A^2,q_1^2,q_2^2)$ and all helicity combinations that do not fulfill $\lambda_1 = \lambda_2 + \lambda_A$ vanish.
Since $\F_1^A(0,0)=0$ due to the crossing property~\eqref{crossing_axial}, these expressions immediately show that the on-shell process $A\to\gamma\gamma$ is forbidden, as stated 
by the Landau--Yang theorem~\cite{Landau:1948kw,Yang:1950rg}. Accordingly, to measure
the differential decay width for the process $A(p,\lambda_A) \to \gamma^*(q_1,\lambda_1) \gamma^*(q_2,\lambda_2)$, given by
\beq
	d\Gamma = \frac{e^4}{32\pi^2} | H_{\lambda_1\lambda_2;\lambda_A} |^2 \frac{\lambda_{A12}^{1/2}}{2m_A^3} d\Omega,
\eeq
one needs at least one virtual photon, with an equivalent two-photon decay width conventionally defined as\footnote{We write everything in decay kinematics, hence for $\tilde\Gamma_{\gamma\gamma}$ to be positive, we use the Minkowskian virtuality $q_1^2>0$.}
\begin{align}
	\tilde\Gamma_{\gamma\gamma} = \lim_{q_1^2\to0} \frac{m_A^2}{q_1^2} \frac{1}{2} \Gamma(A \to \gamma_L^* \gamma_T) .
\end{align}
Averaging over $\lambda_A$ and summing over $\lambda_2=\pm$, we find (the polarization vectors are normalized to one, i.e.\ $\xi_1^2 = q_1^2$):
\begin{align}
	\tilde\Gamma_{\gamma\gamma} = \frac{\pi \alpha^2 m_A}{12}| \F_2^A(0,0) |^2 = \frac{\pi \alpha^2 m_A}{12} | \F_3^A(0,0) |^2.
\end{align}

\subsection[Tensor mesons: $J^{PC} = 2^{++}$]{\boldmath Tensor mesons: $J^{PC} = 2^{++}$}

For the matrix element of a massive tensor meson decaying into two off-shell photons we have
\begin{align}
	\begin{split}
		\< \gamma^*(q_1,\lambda_1) & \gamma^*(q_2,\lambda_2) | T(p,\lambda_T) \> \\
			&= - (2\pi)^4 \delta^{(4)}(q_1+q_2-p) e^2 {\epsilon_\mu^{\lambda_1}}^*(q_1) {\epsilon_\nu^{\lambda_2}}^*(q_2) \int d^4x \, e^{i q_1 \cdot x} \< 0 | T \{ j_\mathrm{em}^\mu(x) j_\mathrm{em}^\nu(0) \} | T(p,\lambda_T) \> \\
			&= i (2\pi)^4 \delta^{(4)}(q_1+q_2-p) e^2 {\epsilon_\mu^{\lambda_1}}^*(q_1) {\epsilon_\nu^{\lambda_2}}^*(q_2) \M^{\mu\nu}(\{p,\lambda_T\} \rightarrow q_1,q_2) \\
			&= i (2\pi)^4 \delta^{(4)}(q_1+q_2-p) e^2 {\epsilon_\mu^{\lambda_1}}^*(q_1) {\epsilon_\nu^{\lambda_2}}^*(q_2) \epsilon_{\alpha\beta}^{\lambda_T}(p) \M^{\mu\nu\alpha\beta}(q_1,q_2),
	\end{split}
\end{align}
with the $T$-matrix elements
\begin{align}
	\M^{\mu\nu}(\{p,\lambda_T\} \rightarrow q_1,q_2) = \epsilon_{\alpha\beta}^{\lambda_T}(p) \M^{\mu\nu\alpha\beta}(q_1,q_2) = i \int d^4x \, e^{i q_1 \cdot x} \< 0 | T \{ j_\mathrm{em}^\mu(x) j_\mathrm{em}^\nu(0) \} | T(p,\lambda_T) \> .
\end{align}
The helicity amplitudes are defined by
\begin{align}
	H_{\lambda_1\lambda_2;\lambda_T} = {\epsilon_\mu^{\lambda_1}}^*(q_1) {\epsilon_\nu^{\lambda_2}}^*(q_2) \epsilon_{\alpha\beta}^{\lambda_T}(p) \M^{\mu\nu\alpha\beta}(q_1,q_2)
\end{align}
and the polarization tensor $\epsilon_{\alpha\beta}^{\lambda_T}$ is constructed as~\cite{Poppe:1986dq}
\begin{align}
	\begin{split}
		\epsilon_{\alpha\beta}^{\pm2}(p) &= \epsilon_\alpha^\pm(p) \epsilon_\beta^\pm(p) , \\
		\epsilon_{\alpha\beta}^{\pm1}(p) &= \frac{1}{\sqrt{2}} \left( \epsilon_\alpha^\pm(p) \epsilon_\beta^0(p) +  \epsilon_\alpha^0(p) \epsilon_\beta^\pm(p) \right) , \\
		\epsilon_{\alpha\beta}^{0}(p) &= \frac{1}{\sqrt{6}} \left( 2 \epsilon_\alpha^0(p) \epsilon_\beta^0(p) + \epsilon_\alpha^+(p) \epsilon_\beta^-(p) + \epsilon_\alpha^-(p) \epsilon_\beta^+(p) \right) ,
	\end{split}
\end{align}
where the polarization vectors are the same as in~\eqref{eq:PolarizationVectorsAxial}. The polarization sum is given by
\begin{align}
	\label{eq:TensorPolarizationSum}
	s^T_{\alpha\beta\alpha'\beta'}(p) := \sum_{\lambda_T} \epsilon_{\alpha\beta}^{\lambda_T}(p) \epsilon_{\alpha'\beta'}^{\lambda_T}(p)^* = \frac{1}{2} \left( s_{\alpha\beta'} s_{\alpha'\beta} + s_{\alpha\alpha'} s_{\beta\beta'} \right) - \frac{1}{3} s_{\alpha\beta} s_{\alpha'\beta'} ,
\end{align}
where
\begin{align}
	s_{\alpha\alpha'} := - \left( g_{\alpha\alpha'} - \frac{p_\alpha p_{\alpha'}}{m_T^2} \right) .
\end{align}
It satisfies
\begin{align}
	g^{\alpha'\alpha''} g^{\beta'\beta''} s^T_{\alpha\beta\alpha'\beta'} s^T_{\alpha''\beta''\alpha'''\beta'''}  =  s^T_{\alpha\beta\alpha'''\beta'''} .
\end{align}

Crossing symmetry requires
\begin{align}
	\M^{\mu\nu\alpha\beta}(q_1,q_2) = \M^{\nu\mu\alpha\beta}(q_2,q_1) .
\end{align}
Furthermore, only those structures can contribute to observables that do not vanish upon contraction with the projector $s^T_{\alpha\beta\alpha'\beta'}$. In particular they have to be symmetric in $\alpha\leftrightarrow\beta$. As for the scalar case, parity conservation excludes the presence of structures with an epsilon tensor, hence the elementary building blocks are again $g^{\mu\nu}$, $q_1^\mu$, $q_2^\mu$.

The BTT construction leads to 20 structures: 7 structures are odd in $\alpha\leftrightarrow\beta$ and 8 more structures vanish upon contraction with the tensor meson projector. Therefore, only five structures contribute to observables:
\begin{align}
	\begin{split}
		T_1^{\mu\nu\alpha\beta} &= g^{\mu\alpha} P_{21}^{\nu\beta} + g^{\nu\alpha} P_{12}^{\mu\beta} + g^{\mu\beta} P_{21}^{\nu\alpha} + g^{\nu\beta} P_{12}^{\mu\alpha} + g^{\mu\nu} (q_1^\alpha q_2^\beta + q_2^\alpha q_1^\beta)  - q_1 \cdot q_2 ( g^{\mu\alpha} g^{\nu\beta} + g^{\nu\alpha} g^{\mu\beta} )  , \\
		T_2^{\mu\nu\alpha\beta} &= (q_1^\alpha q_1^\beta + q_2^\alpha q_2^\beta ) P_{12}^{\mu\nu} , \\
		T_3^{\mu\nu\alpha\beta} &= P_{11}^{\mu\alpha} P_{22}^{\nu\beta} + P_{11}^{\mu\beta} P_{22}^{\nu\alpha} , \\
		T_4^{\mu\nu\alpha\beta} &= P_{12}^{\mu\alpha} P_{22}^{\nu\beta} + P_{12}^{\mu\beta} P_{22}^{\nu\alpha} , \\
		T_5^{\mu\nu\alpha\beta} &= P_{21}^{\nu\alpha} P_{11}^{\mu\beta} + P_{21}^{\nu\beta} P_{11}^{\mu\alpha} ,
	\end{split}
	\label{structures_tensors}
\end{align}
where
\begin{align}
	P_{ij}^{\mu\nu} := g^{\mu\nu} q_i \cdot q_j - q_i^\nu q_j^\mu .
\end{align}
Under photon crossing, these Lorentz structures transform as
\begin{align}
	\Cr{12}{T_{1,2,3}^{\mu\nu\alpha\beta}} = T_{1,2,3}^{\mu\nu\alpha\beta} , \quad \Cr{12}{T_4^{\mu\nu\alpha\beta}} = T_5^{\mu\nu\alpha\beta} .
\end{align}
We define dimensionless form factors $\F_i$, which are the scalar functions in the Lorentz decomposition of the amplitude:
\begin{align}
	\M^{\mu\nu\alpha\beta} = \sum_{i=1}^5 T_i^{\mu\nu\alpha\beta} \frac{1}{m_T^{n_i}} \F_i^T(q_1^2, q_2^2) ,
\end{align}
where $n_1 = 1$ and the other $n_i = 3$.

In terms of these form factors, the helicity amplitudes are
\begin{align}
	\begin{split}
		H_{++;0} = H_{--;0} &= \frac{(q_1^2-q_2^2)^2 - m_T^2(q_1^2+q_2^2)}{\sqrt{6} m_T^3} \F_1^T - \frac{\lambda_{T12}(m_T^2-q_1^2-q_2^2)}{2\sqrt{6} m_T^5} \F_2^T - \sqrt{\frac{2}{3}} \frac{q_1^2 q_2^2}{m_T^3} \F_3^T \\
			&\quad  - \frac{q_2^2(m_T^2-q_1^2-q_2^2)}{\sqrt{6} m_T^3} \F_4^T   - \frac{q_1^2(m_T^2-q_1^2-q_2^2)}{\sqrt{6} m_T^3} \F_5^T  , \\
		H_{+-;+2} = H_{-+;-2} &= - \frac{m_T^2-q_1^2-q_2^2}{m_T} \F_1^T - \frac{2q_1^2q_2^2}{m_T^3} \F_3^T  - \frac{q_2^2(m_T^2-q_1^2-q_2^2)}{m_T^3} \F_4^T   - \frac{q_1^2(m_T^2-q_1^2-q_2^2)}{m_T^3} \F_5^T  , \\
		H_{+0;+1} = H_{-0;-1} &= \frac{q_2^2}{\xi_2} \bigg( \frac{m_T^2+q_1^2-q_2^2}{\sqrt{2} m_T^2} \F_1^T + \frac{q_1^2 (m_T^2 - q_1^2 + q_2^2)}{\sqrt{2} m_T^4} \F_3^T \\
			&\quad + \frac{(m_T^2-q_1^2-q_2^2)(m_T^2-q_1^2+q_2^2)}{2\sqrt{2} m_T^4} \F_4^T + \frac{q_1^2 (m_T^2+q_1^2-q_2^2)}{\sqrt{2} m_T^4} \F_5^T \bigg) , \\
		H_{0+;-1} = H_{0-;+1} &= - \frac{q_1^2}{\xi_1} \bigg( \frac{m_T^2-q_1^2+q_2^2}{\sqrt{2} m_T^2} \F_1^T + \frac{q_2^2 (m_T^2 + q_1^2 - q_2^2)}{\sqrt{2} m_T^4} \F_3^T \\
			&\qquad\quad + \frac{q_2^2(m_T^2-q_1^2+q_2^2)}{\sqrt{2} m_T^4} \F_4^T + \frac{(m_T^2-q_1^2-q_2^2)(m_T^2+q_1^2-q_2^2)}{2\sqrt{2} m_T^4} \F_5^T \bigg) , \\
		H_{00;0} &= \frac{q_1^2q_2^2}{\xi_1\xi_2} \bigg( \sqrt{\frac{2}{3}} \frac{2}{m_T} \F_1^T - \frac{\lambda_{T12}}{\sqrt{6} m_T^5} \F_2^T + \frac{m_T^4-(q_1^2-q_2^2)^2}{\sqrt{6} m_T^5} \F_3^T \\*
			&\qquad\quad + \frac{(m_T^2-q_1^2+q_2^2)^2}{\sqrt{6}m_T^5} \F_4^T + \frac{(m_T^2+q_1^2-q_2^2)^2}{\sqrt{6}m_T^5} \F_5^T \bigg) ,
	\end{split}
	\label{hel_tensor}
\end{align}
where $\lambda_{T12} := \lambda(m_T^2,q_1^2,q_2^2)$ and again only amplitudes fulfilling $\lambda_1 = \lambda_2 + \lambda_T$ do not vanish.

Finally, the differential decay width for the process $T(p,\lambda_T) \to \gamma^*(q_1,\lambda_1) \gamma^*(q_2,\lambda_2)$ is given by
\begin{align}
	d\Gamma = \frac{e^4}{32\pi^2} | H_{\lambda_1\lambda_2;\lambda_T} |^2 \frac{\lambda_{T12}^{1/2}}{2m_T^3} d\Omega,
\end{align}
leading to the on-shell result
\begin{align}
\label{onshell_tensor}
	\Gamma_{\gamma\gamma} &= \frac{\pi \alpha^2 m_T}{5} \left( | \F_1^T(0,0) |^2  + \frac{1}{24} | \F_2^T(0,0) |^2 \right) .
\end{align}


\section{Brodsky--Lepage limit for the transition form factors}
\label{sec:BL}

\subsection{Pseudoscalar mesons}

We start again with a review of the familiar pseudoscalar case~\cite{Lepage:1979zb,Lepage:1980fj}, restricting the analysis to the leading-order result. In addition to the definition of the TFF~\eqref{TFF_pion}
we need the decay constants $F_P^a$
\beq
\langle 0|\bar q(0) \gamma_\mu\gamma_5\frac{\lambda^a}{2} q(0)|P(p)\rangle = i p_\mu F_P^a,
\eeq
with flavor decomposition using the Gell-Mann matrices $\lambda_a$ and $\lambda_0=\sqrt{2/3}\,\mathds{1}$.
The wave functions $\phi_P^a(u)$ are then defined as
\beq
\langle 0|\bar q(x) \gamma_\mu\gamma_5\frac{\lambda^a}{2} q(0)|P(p)\rangle = i p_\mu F_P^a
\int_0^1\diff u\, e^{-iu p\cdot x}\phi_P^a(u),
\label{wave_function}
\eeq
where the path-ordered gauge factor to connect the quark fields at points $0$ and $x$ on the left-hand side has been omitted~\cite{Balitsky:1987bk}.
Asymptotically, the wave functions can be calculated based on conformal symmetry of QCD (see~\cite{Braun:2003rp} for a review), with the result
\beq
\phi_P^a(u)=6u(1-u)\equiv \phi(u).
\label{phi_asym}
\eeq
For all TFFs, we will only consider asymptotic results, and to the extent possible we will write the corresponding wave functions in terms of $\phi(u)$ as it appears in the pseudoscalar case. 
Beyond the asymptotic result, the matrix element in~\eqref{wave_function} and thus the wave function become scale dependent, but the conformal analysis shows that the higher-order terms can be organized in an expansion in Gegenbauer polynomials $C_n^{3/2}$, 
\beq
\phi(u,\mu)=6u(1-u)\sum_{n=0}^\infty a_n(\mu)C_n^{3/2}(2u-1),
\label{wave_function_expansion}
\eeq
with $a_0=1$ and the scale dependence, affecting the coefficients with $n>1$, determined by
\beq
a_n(\mu)=a_n(\mu_0)\bigg(\frac{\alpha_s(\mu)}{\alpha_s(\mu_0)}\bigg)^{\gamma_n^{(0)}/\beta_0},
\eeq
where
\beq
\gamma_n^{(0)}=C_F\bigg(1-\frac{2}{(n+1)(n+2)}+4\sum_{m=2}^{n+1}\frac{1}{m}\bigg),\qquad \beta_0=\frac{11}{3}N_c-\frac{2}{3}N_f,\qquad C_F=\frac{N_c^2-1}{2N_c}.
\eeq
Due to $C_0^{3/2}=1$ and the orthogonality relation
\beq
\int_0^1\diff u\, u(1-u)C_n^{3/2}(2u-1)C_m^{3/2}(2u-1)
=\delta_{nm}\frac{(n+1)(n+2)}{4(2n+3)},
\eeq
the expansion~\eqref{wave_function_expansion} automatically fulfills the normalization condition
\beq
\int_0^1\diff u\, \phi(u,\mu)=1.
\eeq
Further, charge-conjugation and translation invariance imply $\phi_P^a(u) = \eta(a) \phi_{P^c}^a(1-u)$, with $\eta(a) = +1$ for $a\in\{0,1,3,4,6,8\}$ and $\eta(a)=-1$ for $a\in\{2,5,7\}$, and where $P^c$ denotes the $C$ conjugate of $P$. In particular, for $P=P^c$ and $a\in\{0,1,3,4,6,8\}$ the odd coefficients in the Gegenbauer expansion vanish.

The leading diagrams in the BL formalism are obtained from contracting the quark fields in the time-ordered product using free propagators, which leads to
\beq
 T\{ j^{\mu}_\mathrm{em}(x)j^{\nu}_\mathrm{em}(0) \}=\bar q(x)\Q^2\gamma^\mu\gamma^\alpha\gamma^\nu q(0) S^F_\alpha(x)
 +\bar q(0) \Q^2 \gamma^\nu\gamma^\alpha\gamma^\mu q(x) S^F_\alpha(-x),
\eeq
where 
\beq
\label{SF}
S^F_\mu(x)=i\int\frac{\diff^4 p}{(2\pi)^4}\frac{p_\mu e^{-ip\cdot x}}{p^2+i\eps}=\frac{i x_\mu}{2\pi^2(x^2-i\eps)^2}.
\eeq
The remaining Dirac structure becomes
\beq
\gamma^\mu\gamma^\alpha\gamma^\nu=g^{\mu\alpha}\gamma^\nu+g^{\nu\alpha}\gamma^\mu-g^{\mu\nu}\gamma^\alpha+i\eps^{\mu\alpha\nu\beta}\gamma_\beta \gamma_5.
\eeq
Using translational invariance and the symmetry of the wave function under $u\to 1-u$, both contractions yield the same result, and since the matrix element of the vector current vanishes, this leads to
\begin{align}
 \M_{\mu\nu}&=
 i\int \diff^4 x e^{iq_1 \cdot x}(2i\eps_{\mu\alpha\nu\beta})\langle 0|\bar q(x)\Q^2\gamma^\beta\gamma_5 q(0)|P(p)\rangle S_F^\alpha(x)\notag\\
 &=-4i\sum_a C_a F_P^a \eps_{\mu\alpha\nu\beta} (q_1+q_2)^\beta \int_0^1\diff u\, \phi(u)\int \diff^4 x e^{iq_1 \cdot x} e^{-iu p\cdot x} S_F^\alpha(x),
\end{align}
with flavor weights $C_a=\frac{1}{2}\Tr(\Q^2\lambda^a)$, i.e.,
\beq
\label{weights}
C_3=\frac{1}{6},\qquad C_8=\frac{1}{6\sqrt{3}},\qquad C_0=\frac{2}{3\sqrt{6}}.
\eeq
The Feynman propagator fulfills the relations
\begin{align}
\int\diff^4 x\,S_F^\mu(x) e^{i q\cdot x}&=i\frac{q^\mu}{q^2},\qquad
\int\diff^4 x\, x^\mu S_F^\nu(x)e^{i q\cdot x}=\frac{g^{\mu\nu}}{q^2}-\frac{2q^\mu q^\nu}{q^4},\notag\\
\int\diff^4 x\, x^\mu x^\nu S_F^\lambda(x)e^{i q\cdot x}&=\frac{2i}{q^4}\bigg(g^{\mu\nu} q^\lambda+g^{\mu\lambda} q^\nu+g^{\nu\lambda} q^\mu-\frac{4q^\mu q^\nu q^\lambda}{q^2}\bigg),
\end{align}
leading to ($q=q_1-u p$)
\begin{align}
 \M_{\mu\nu}&=4\sum_a C_a F_P^a \eps_{\mu\alpha\nu\beta} (q_1+q_2)^\beta \int_0^1\diff u\, \phi(u)\,\frac{q^\alpha}{q^2}\notag\\
 &=-4\sum_a C_a F_P^a \eps_{\mu\nu\alpha\beta}q_1^{\alpha}q_2^{\beta} \int_0^1\diff u\,\frac{\phi(u)}{(1-u)q_1^2+uq_2^2-u(1-u)m_P^2}.
\end{align}
Reading off the result for the TFF,
\beq
\label{pion_TFF_result}
F_{P\gamma^*\gamma^*}(q_1^2,q_2^2)=-4\sum_a C_a F_P^a
\int_0^1\diff u\,\frac{\phi(u)}{u q_1^2+(1-u)q_2^2-u(1-u)m_P^2},
\eeq
this reproduces the expected asymptotic behavior~\eqref{eq:pQCD}.

We stress that while we have kept the mass $m_P$ in the final result, this leading-order derivation does not provide a consistent treatment of mass effects. To this end, one would have to differentiate between the meson momentum $p$ and the light-cone momentum 
\beq
k_\mu= p_\mu - x_\mu \frac{m_P^2}{2p\cdot x},
\eeq
which would appear in the exponential in~\eqref{wave_function}. Accordingly, including terms of $\Order(m_P^2)$ would require the consideration of subleading terms in the light-cone expansion. Moreover, we stress that the result~\eqref{pion_TFF_result} can only be strictly justified from an OPE in the limit in which both photon virtualities are large, otherwise, the wave function approach amounts to a resummation of higher-order terms in the OPE~\cite{Manohar:1990hu}.
This BL factorization into a non-perturbative wave function and a perturbatively calculable kernel can be derived in soft-collinear effective theory (SCET)~\cite{Bauer:2002nz,Rothstein:2003wh}, see also~\cite{Grossmann:2015lea}. 
In this language, the SCET Wilson coefficient is calculable in perturbation theory and the pion wave function becomes the matrix element of a SCET operator.

\subsection{Scalar mesons}

For the scalar mesons we largely follow the definition of the wave functions from~\cite{Cheng:2005nb,Lu:2006fr}. First, in general, the decay constant can be equivalently defined for the vector or the scalar current
\begin{align}
\langle 0|\bar q(0) \gamma_\mu\frac{\lambda^a}{2} q(0)|S(p)\rangle &=- p_\mu F_S^a,\notag\\
\langle 0|\bar q(0) \frac{\lambda^a}{2}  q(0)|S(p)\rangle &=m_S \bar F_S^a(\mu),
\end{align}
related by the conservation of the vector current according to
\beq
F_S^a=if^{abc}\bar F_S^b(\mu) \frac{\Tr(\M \lambda^c)}{m_S},\qquad \M=\text{diag}\big(m_u,m_d,m_s\big),
\eeq
where the scale dependence in $\bar F_S^a(\mu)$ is canceled by the one of the quark masses.  
However, for $a=0,3,8$ this implies $F_S^a=0$, in such a way that the leading term in the light-cone expansion vanishes. In fact, contrary to the pseudoscalar mesons, only odd powers in the Gegenbauer expansion contribute, where the normalization 
\beq
\int_0^1\diff u\, \phi_S^a(u,\mu)=0
\eeq
reflects the fact that $F_S^a=0$. Therefore, the first non-vanishing term involves an unknown Gegenbauer coefficient, which could be made dimensionless by factoring out the scalar decay constant $\bar F_S^a$. Following the notation in the literature~\cite{Cheng:2005nb,Lu:2006fr,Kroll:2016mbt} we write 
\beq
\langle 0|\bar q(x) \gamma_\mu\frac{\lambda^a}{2} q(0)|S(p)\rangle = -p_\mu \bar F_S^a(\mu) B_1(\mu)
\int_0^1\diff u\, e^{-iu p\cdot x}3(2u-1)\phi(u),
\eeq
where $B_1(\mu)$ refers to the Gegenbauer coefficient (assuming that all the flavor dependence is captured by $\bar F_S^a(\mu)$). 
In close analogy to the calculation for the pion TFF this leads to
\beq
\M_{\mu\nu}=4\sum_a C_a \bar F_S^a(\mu) B_1(\mu)\int_0^1\diff u\,\frac{3(2u-1)\phi(u)}{q^2}\big(q_\mu p_\nu+q_\nu p_\mu-g_{\mu\nu} p\cdot q\big),
\eeq
where again $q=q_1-u p$.
In contrast to the pseudoscalar case this expression is only manifestly gauge invariant for $m_S=0$. In this limit, direct projection onto the BTT structures produces a singularity in $\F_2^S$ at $q_1\cdot q_2 = 0$, which, however, is only apparent. It can be removed using $m_S^2 = q_1^2 + 2 q_1 \cdot q_2 + q_2^2 = 0$ and integration by parts. This leads to our final result for the scalar TFFs:
\begin{align}
 \F_1^S(q_1^2,q_2^2)&=4\sum_a C_a \bar F_S^a(\mu) B_1(\mu) m_S\int_0^1\diff u\,\frac{3(2u-1)^2\phi(u)}{u q_1^2+(1-u)q_2^2},\notag\\
 \F_2^S(q_1^2,q_2^2) 
 	&= 4 \sum_a C_a \bar F_S^a(\mu) B_1(\mu) m_S^3 \int_0^1\diff u\, \frac{3u(1-u)\phi(u)}{( u q_1^2+(1-u)q_2^2)^2} .
 \label{result_scalar}
\end{align}

\subsection{Axial-vector mesons}

We will use the axial-vector distribution amplitudes from~\cite{Yang:2005gk,Yang:2007zt}, which are derived in close analogy to the vector-meson case~\cite{Ball:1998sk,Ball:1998ff}. First, the decay constants are defined as 
\beq
 \langle 0|\bar q(0) \gamma_\mu\gamma_5\frac{\lambda^a}{2} q(0)|A(p,\lambda_A)\rangle 
 = F_A^a m_A \eps_\mu.
\eeq
The main complication compared to the (pseudo-) scalar mesons is that the polarization vector contributes to different orders in the twist expansion, so that, at each order, a different wave function may occur. The different orders are separated by defining a light-cone vector 
\beq
k_\mu= p_\mu - x_\mu \frac{m_A^2}{2p\cdot x},
\eeq
which on the light cone $x^2=0$ fulfills $k^2=0$. The polarization vector is then decomposed according to
\beq
\eps^\mu = \frac{\eps\cdot x}{k\cdot x} k^\mu + \frac{\eps\cdot k}{k\cdot x} x^\mu +\eps^\mu_\perp
= \frac{\eps\cdot x}{k\cdot x} \bigg(k^\mu -\frac{m_A^2}{2k\cdot x}x^\mu\bigg) +\eps^\mu_\perp,
\eeq
because due to $p\cdot\eps=0$ one has $\eps\cdot k=-\eps\cdot x\frac{m_A^2}{2k\cdot x}$.
This decomposition gives rise to three different wave functions occurring in the axial-vector matrix element
\beq
\label{axial_matrix_element}
\langle 0|\bar q(x) \gamma^\mu\gamma_5\frac{\lambda^a}{2} q(0)|A(p,\lambda)\rangle 
= F_A^a m_A
\int_0^1\diff u\, e^{-iu k\cdot x}\bigg[k^\mu \frac{\eps\cdot x}{k\cdot x}\phi(u)
+ \eps^\mu_\perp \phi_{\perp}(u)-x^\mu \frac{m_A^2\,\eps\cdot x}{2(k\cdot x)^2} \phi_{3}(u)\bigg].
\eeq
Here, $\phi_{\perp}(u)$ and $\phi_{3}(u)$ are of higher twist. To obtain a gauge-invariant result for the TFFs, these wave functions should be replaced by so-called Wandzura--Wilczek relations~\cite{Wandzura:1977qf} in terms of the leading twist-$2$ distribution amplitudes, which effectively neglects three-parton contributions. In this approximation we have~\cite{Yang:2007zt}
\beq
\label{WW_axial}
\phi_{\perp}(u)=\frac{1}{2}\bigg(\int_0^u\diff v\frac{\phi(v)}{1-v}+\int_u^1\diff v\frac{\phi(v)}{v}\bigg)
=\frac{1}{2}\big(3-\phi(u)\big)
\eeq
for the asymptotic $\phi(u)$ from~\eqref{phi_asym}, while $\phi_{3}(u)$ does not actually contribute due to the antisymmetry of the $\epsilon$ tensor, but could be obtained with similar methods from~\cite{Ball:1998ff}.
In contrast to the pseudoscalar case, there is now also a non-vanishing contribution from the vector matrix element
\beq
\label{A_vector}
\langle 0|\bar q(x) \gamma^\mu\frac{\lambda^a}{2} q(0)|A(p,\lambda_A)\rangle 
= -\frac{1}{4}F_A^a m_A \eps^{\mu\nu\alpha\beta}\eps_\nu k_\alpha x_\beta
\int_0^1\diff u\, e^{-iu k\cdot x}
\phi(u).
\eeq
This is again a twist-$3$ contribution, which technically requires another wave function, but in the same approximation as~\eqref{WW_axial} this new wave function becomes
\beq
2(1-u)\int_0^u\diff v\frac{\phi(v)}{1-v}+2u\int_u^1\diff v\frac{\phi(v)}{v}
=\phi(u)
\eeq
asymptotically.

Starting from
\begin{align}
\eps_\alpha M^{\mu\nu\alpha}&=4i\sum_a C_a\int\diff^4 x\, e^{iq_1 \cdot x}
\bigg(i\eps^{\mu\alpha\nu\beta} \langle 0|\bar q(x) \gamma_\beta\gamma_5\frac{\lambda^a}{2} q(0)|A(p,\lambda_A)\rangle\notag\\
&+\langle 0|\bar q(x) \big(g^{\mu\alpha}\gamma^\nu
+g^{\nu\alpha}\gamma^\mu-g^{\mu\nu}\gamma^\alpha\big)\frac{\lambda^a}{2} q(0)|A(p,\lambda_A)\rangle\bigg)S^F_{\alpha}(x),
\end{align}
the decomposition of the vector and axial-vector matrix elements~\eqref{axial_matrix_element} and~\eqref{A_vector} gives 
\begin{align}
\label{int_result}
\eps_\alpha M^{\mu\nu\alpha}&=4i\sum_a C_a F_A^a m_A\int_0^1\diff u \int\diff^4 x e^{iq\cdot x}\bigg[i\eps^{\mu\nu\beta\alpha}S^F_{\alpha}(x)\Big(p_\beta \frac{\eps\cdot x}{p\cdot x}\big(\phi(u)-\phi_\perp(u)\big)+\eps_\beta \phi_{\perp}(u)\Big)\notag\\
&-\frac{1}{4}\eps^{\nu\alpha\beta\gamma}\eps_\alpha p_\beta x_\gamma S_F^\mu(x) \phi(u)
-\frac{1}{4}\eps^{\mu\alpha\beta\gamma}\eps_\alpha p_\beta x_\gamma S_F^\nu(x) \phi(u)\bigg],
\end{align}
where we have again neglected higher terms in the light-cone expansion. 
To perform the integral, we define
\beq
\Phi(u)=\int_0^u\diff v\Big(\phi(v)-\phi_{\perp}(v)\Big)
=\frac{2u-1}{4}\phi(u)
\eeq
and integrate by parts to obtain
\begin{align}
 \int_0^1\diff u \int\diff^4 x e^{iq\cdot x}S_{F}^\mu(x)  \frac{x^\nu}{p\cdot x}\big(\phi(u)-\phi_{\perp}(u)\big)
 &=i \int_0^1\diff u \int\diff^4 x e^{iq\cdot x}S_{F}^\mu(x) x^\nu\Phi(u)\notag\\
 &=i \int_0^1\diff u\, \Phi(u)\bigg(\frac{g^{\mu\nu}}{q^2}-\frac{2q^\mu q^\nu}{q^4}\bigg).
\end{align}

The integrals in~\eqref{int_result} become
\begin{align}
\label{result_raw}
\eps_\alpha M^{\mu\nu\alpha}&=4i\sum_a C_a F_A^a m_A \eps_\alpha\int_0^1\diff u 
\bigg[\Phi(u) \Big(\eps^{\alpha\mu\nu\beta}\frac{p_\beta}{q^2}
-\frac{1}{q^4}\eps^{\mu\nu\beta\gamma}(q_1-q_2)^\alpha q_{1\beta} q_{2\gamma}\Big)\notag\\
&
-\eps^{\alpha\mu\nu\beta}\frac{q_\beta}{q^2} \phi_{\perp}(u)+\frac{1}{2q^4} \phi(u)\Big(\eps^{\alpha\mu\beta\gamma}q^\nu q_{1\beta} q_{2\gamma}
+\eps^{\alpha\nu\beta\gamma}q^\mu q_{1\beta} q_{2\gamma}\Big)\bigg].
\end{align}
This expression is already manifestly gauge invariant even for non-zero $m_A$:
\begin{align}
	q_{1\mu} \epsilon_\alpha M^{\mu\nu\alpha} &= 4i\sum_a C_a F_A^a m_A \eps_\alpha \eps^{\alpha\mu\nu\beta} q_{1\mu} q_{2\beta} \int_0^1\diff u 
\frac{1}{q^4} \bigg[ q^2 \Big( \Phi(u) + u \phi_{\perp}(u) \Big)
-\frac{q_1 \cdot q}{2} \phi(u)\bigg] \nn
	&= 4i\sum_a C_a F_A^a m_A \eps_\alpha \eps^{\alpha\mu\nu\beta} q_{1\mu} q_{2\beta} \int_0^1\diff u \frac{\partial}{\partial u} \left( \frac{3u^2(u-1)}{2 q^2} \right) = 0.
\end{align}
Therefore, the form factors~\eqref{eq:AxialFormFactors} can be obtained directly by projecting onto the BTT decomposition. The projectors following from the BTT derivation in Sect.~\ref{sec:BTTAxials} lead to spurious divergences in $\frac{1}{q_1 \cdot q_2}$, which, however, can be shown to vanish by expressing all scalar products in terms of $q_1\cdot q_2$, $q^2$, and $\frac{\partial}{\partial u} \frac{1}{q^2}$, as well as integration by parts. This leads to the following results for the axial-vector TFFs:
\begin{align}
	\label{result_final}
	\F_1^A(q_1^2,q_2^2)&=\Order\big(q_i^{-6}\big),\nn
	\F_2^A(q_1^2,q_2^2)&=4\sum_a C_a F_A^a m_A^3 \,\int_0^1 \diff u \frac{u\,\phi(u)}{(u q_1^2 +(1-u)q_2^2-u(1-u)m_A^2)^2},\nn
	\F_3^A(q_1^2,q_2^2)&=-4\sum_a C_a F_A^a m_A^3 \,\int_0^1 \diff u \frac{(1-u)\phi(u)}{(u q_1^2 +(1-u)q_2^2-u(1-u)m_A^2)^2}.
\end{align}
We checked that the same results are obtained by expressing~\eqref{result_raw} explicitly in terms of the $\bar T_i^{\mu\nu\alpha}$ and reducing the final result by means of the Schouten identities. 
In particular, we find that the contribution to $\F_1$ cancels altogether at this order, and that $\F_2(0,q^2)$ does not converge. 
This logarithmic end-point singularity has been observed before in the context of holographic models of QCD~\cite{Leutgeb:2019gbz}. 
Since~\eqref{result_raw} is gauge invariant and free of kinematic singularities even for finite $m_A$, it is meaningful to keep the axial-vector mass in our final result~\eqref{result_final}, similarly to the pseudoscalar case. 
Finally, we note that the predictions for the helicity amplitudes~\eqref{hel_axial} are not affected by the divergence, since the TFFs contributing in the respective singly-virtual limits are well-behaved.

\subsection{Tensor mesons}

In the same way as for the scalar mesons, the leading-order coupling of tensor mesons to vector and axial-vector currents vanishes, so that again the result of the light-cone analysis would be sensitive to the first Gegenbauer coefficient. This Gegenbauer coefficient is usually replaced in terms of decay constants $F_T^a$ defined as~\cite{Braun:2000cs,Cheng:2010hn}
\beq
\langle 0| j_{\mu\nu}(0) |T(p,\lambda_T) \rangle=F_T^a m_T^2 \eps_{\mu\nu}^{\lambda_T}, \quad j_{\mu\nu}(x) = \bar q(x) \frac{1}{2}\Big(\gamma_\mu i \overset{\leftrightarrow}{D}_\nu + \gamma_\nu i \overset{\leftrightarrow}{D}_\mu\Big)\frac{\lambda^a}{2} q(x),
\eeq
with covariant derivative $\overset{\leftrightarrow}{D}_\mu=D_\mu - \overset{\leftarrow}{D}_\mu$. 
In terms of these decay constants the expressions for the matrix elements become~\cite{Braun:2000cs,Cheng:2010hn}
\begin{align}
\langle 0|\bar q(x) \gamma^\mu\gamma_5\frac{\lambda^a}{2} q(0)|T(p,\lambda_T)\rangle 
&= F_T^a m_T^2 \eps^{\mu\nu\alpha\beta}\eps_{\beta\delta} x_\nu \frac{k_\alpha x^\delta}{2k\cdot x}
\int_0^1\diff u\, e^{-iu k\cdot x}
\phi_a(u),\notag\\
\langle 0|\bar q(x) \gamma^\mu\frac{\lambda^a}{2} q(0)|T(p,\lambda_T)\rangle 
&= F_T^a m_T^2
\int_0^1\diff u\, e^{-iu k\cdot x}\notag\\
&\times\bigg[k^\mu \frac{\eps_{\alpha\beta} x^\alpha x^\beta}{(k\cdot x)^2}\phi_1(u)
+ \frac{\eps^{\mu\alpha}_\perp x_\alpha}{k\cdot x} \phi_2(u)-x^\mu \frac{\eps_{\alpha\beta} x^\alpha x^\beta}{2(k\cdot x)^3} m_T^2\phi_{3}(u)\bigg],
\end{align}
with asymptotic wave functions
\beq
\phi_1(u)=5(2u-1)\phi(u),\qquad \phi_2(u)=5(2u-1)^3,\qquad
 \phi_a(u)=\frac{1}{3}\phi_1(u),
\eeq
and 
\beq
\eps^{\alpha\beta}_\perp x_\beta = \eps^{\alpha\beta} x_\beta - \frac{\eps^{\beta\gamma} x_\beta x_\gamma}{k\cdot x}\bigg(k^\alpha -\frac{m_T^2}{2k\cdot x} x^\alpha\bigg).
\eeq
As before, we do not keep subleading terms in the light-cone expansion, including terms proportional to $m_T^2$, so that again $\phi_3(u)$ does not play a role. 

Removing the poles in $k\cdot x=p\cdot x$ using the same strategy as for the axial-vector case, we obtain as intermediate result
\begin{align}
\label{result_tensor}
 \M^{\mu\nu\alpha\beta}&=4\sum_a C_a F_T^a m_T^2\int_0^1\diff u\, \frac{5}{6}\phi(u)\bigg[\frac{1-2u(1-u)}{q^2}\big(g^{\mu\alpha}g^{\nu\beta}+g^{\mu\beta}g^{\nu\alpha}\big)+\frac{6u(1-u)}{q^4}g^{\mu\nu}q^\alpha q^\beta\notag\\
 &+\frac{u}{q^4}\big((4u^2-5u+1)q_1^\nu+(4u^2-3u+1)q_2^\nu\big)\big(g^{\mu\alpha}q^\beta+g^{\mu\beta}q^\alpha\big)\notag\\
 &-\frac{1-u}{q^4}\big((4u^2-5u+2)q_1^\mu+u(4u-3)q_2^\mu\big)\big(g^{\nu\alpha}q^\beta+g^{\nu\beta}q^\alpha\big)\\
 &+\frac{8u(1-u)q^\alpha q^\beta}{q^6}\Big((2u-1)\big((1-u)q_1^\mu q_1^\nu-u q_2^\mu q_2^\nu\big)-(1-2u(1-u))q_1^\mu q_2^\nu+2u(1-u)q_2^\mu q_1^\nu\Big)\notag
 \bigg],
\end{align}
where we have already dropped terms involving $g^{\alpha\beta}$ because they cancel upon contraction with the (trace-free) polarization tensor.
The expression~\eqref{result_tensor} is not manifestly gauge invariant yet, as the contraction with $q_1^\mu$ only vanishes up to terms that disappear after contraction with the polarization tensor. To remove these unphysical terms we apply projectors onto the five relevant structures~\eqref{structures_tensors}, which allows us to identify  
\begin{align}
\label{result_final_tensor}
 \F_1^T(q_1^2,q_2^2)
 &=4\sum_a C_a F_T^a m_T^3 \,\int_0^1 \diff u \,\phi(u)\frac{5u(1-u)(3-20u(1-u))}{6(u q_1^2 +(1-u)q_2^2)^2},\notag\\
\F_2^T(q_1^2,q_2^2)
&=-4\sum_a C_a F_T^a m_T^5 \,\int_0^1 \diff u \,\phi(u)\frac{20u^2(1-u)^2}{3(u q_1^2 +(1-u)q_2^2)^3},\notag\\
\F_3^T(q_1^2,q_2^2)
&=4\sum_a C_a F_T^a m_T^5 \,\int_0^1 \diff u \,\phi(u)\frac{10u(1-u)(1-2u(1-u))}{3(u q_1^2 +(1-u)q_2^2)^3},\notag\\
\F_4^T(q_1^2,q_2^2)
&=-4\sum_a C_a F_T^a m_T^5 \,\int_0^1 \diff u \,\phi(u)\frac{10(2u-1)u(1-u)^2}{3(u q_1^2 +(1-u)q_2^2)^3},\notag\\
\F_5^T(q_1^2,q_2^2)
&=4\sum_a C_a F_T^a m_T^5 \,\int_0^1 \diff u \,\phi(u)\frac{10(2u-1)u^2(1-u)}{3(u q_1^2 +(1-u)q_2^2)^3}.
\end{align}
As in the case of the scalar meson, the kinematic singularities at $q_1\cdot q_2$ indeed cancel, but only as long as $m_T=0$. For that reason, our calculation again does not capture terms $\Order(m_T^2)$ consistently. Similarly to the axial-vector case, we find singularities in the singly-virtual limits of
$\F_{3\text{--}5}^T$. However, the helicity amplitudes~\eqref{hel_tensor} are still well-defined even for singly-virtual kinematics, because only TFFs that remain finite in the respective limits contribute.

\subsection{Summary of Brodsky--Lepage scaling}

\begin{figure}[t]
	\centering
	\includegraphics[width=0.6\linewidth]{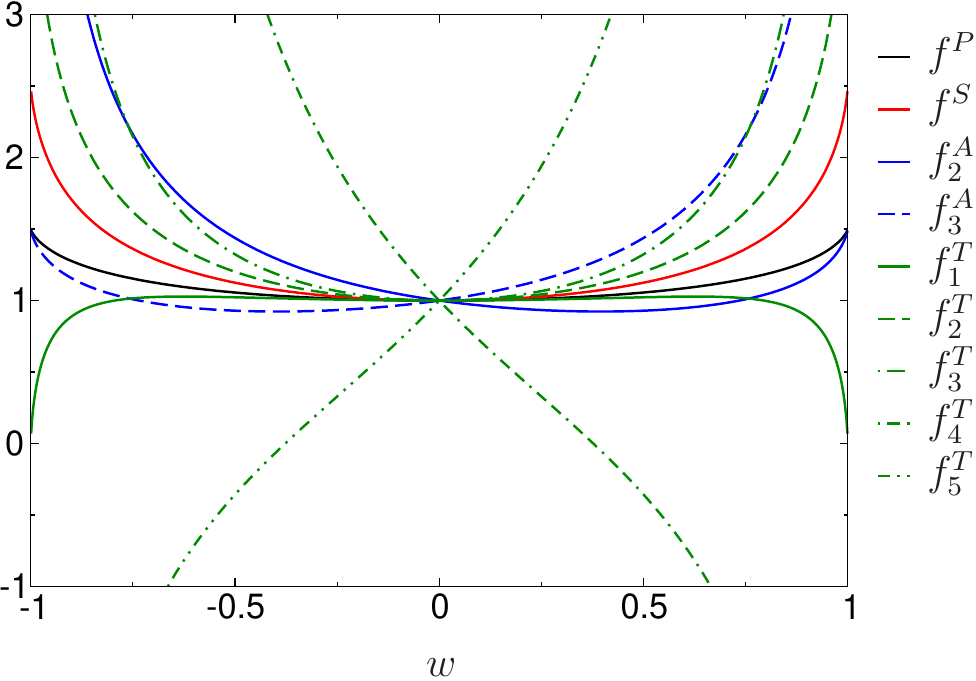}
	\caption{Asymmetry functions for pseudoscalar, scalar, axial-vector, and tensor mesons. All functions are normalized to their value at $w=0$.}      
	\label{fig:fw}
\end{figure}

We summarize our results in terms of their scaling in the average photon virtualities $Q^2$ and the asymmetry parameter $w$
\beq
\label{defQ}
Q^2=\frac{q_1^2+q_2^2}{2},\qquad w=\frac{q_1^2-q_2^2}{q_1^2+q_2^2}.
\eeq
Separating the flavor decomposition and mass factors, we have
\begin{align}
\label{BL_scaling}
 F_{P\gamma^*\gamma^*}(q_1^2,q_2^2)&=\frac{4\sum_a C_a F_P^a}{Q^2}f^P(w),\notag\\
 \F_1^S(q_1^2,q_2^2)&=\frac{4\sum_a C_a \bar F_S^a(\mu) B_1(\mu) m_S}{Q^2}f_1^S(w),\notag\\
 \F_2^S(q_1^2,q_2^2)&=\frac{4\sum_a C_a \bar F_S^a(\mu) B_1(\mu) m_S^3}{Q^4}f_2^S(w),\notag\\
 \F_1^A(q_1^2,q_2^2)&=\Order(Q^{-6}),\notag\\
 \F_i^A(q_1^2,q_2^2)&=\frac{4\sum_a C_a F_A^a m_A^3}{Q^4}f_i^A(w),\qquad i\in\{2,3\},\notag\\
 \F_1^T(q_1^2,q_2^2)&=\frac{4\sum_a C_a F_T^a m_T^3}{Q^4}f_1^T(w),\notag\\
 \F_i^T(q_1^2,q_2^2)&=\frac{4\sum_a C_a F_T^a m_T^5}{Q^6}f_i^T(w),\qquad i\in\{2,3,4,5\},
\end{align}
with asymmetry functions
\begin{align}
 f^P(w)&=-\frac{3}{2w^2}\bigg(1+\frac{1-w^2}{2w}\log\frac{1-w}{1+w}\bigg),\notag\\
 f^S(w)&\equiv f^S_1(w)=f^S_2(w)=\frac{3}{2w^4}\bigg(3-2w^2+3\frac{1-w^2}{2w}\log\frac{1-w}{1+w}\bigg),\notag\\
 f^A_2(w)&=\frac{3}{4w^3}\bigg(3-2w+\frac{(3+w)(1-w)}{2w}\log\frac{1-w}{1+w}\bigg),\notag\\
 f^A_3(w)&=\frac{3}{4w^3}\bigg(3+2w+\frac{(3-w)(1+w)}{2w}\log\frac{1-w}{1+w}\bigg),\notag\\
 f^T_1(w)&=\frac{5(1-w^2)}{8w^6}\bigg(15-4w^2+\frac{3(5-3w^2)}{2w}\log\frac{1-w}{1+w}\bigg),\notag\\
 f^T_2(w)&=-\frac{5}{8w^6}\bigg(15-13w^2+\frac{3(1-w^2)(5-w^2)}{2w}\log\frac{1-w}{1+w}\bigg),\notag\\
 f^T_3(w)&=-\frac{5}{8w^6}\bigg(15-w^2-\frac{w^4+6w^2-15}{2w}\log\frac{1-w}{1+w}\bigg),\notag\\
 f^T_4(w)&=-\frac{5}{24w^6}\bigg(45+30w-21w^2-8w^3+\frac{3(1+w)(15-5w-7w^2+w^3)}{2w}\log\frac{1-w}{1+w}\bigg),\notag\\
 f^T_5(w)&=-\frac{5}{24w^6}\bigg(45-30w-21w^2+8w^3+\frac{3(1-w)(15+5w-7w^2-w^3)}{2w}\log\frac{1-w}{1+w}\bigg).
\end{align}
These functions are shown in Fig.~\ref{fig:fw}, together with their limiting cases in Table~\ref{tab:asymmetry}.

\begin{table}[t]
 \renewcommand{\arraystretch}{1.3} 
\begin{center}
\begin{tabular}{crrr}
\toprule
$w$ &$+1$& $0$&$-1$ \\ \midrule
$f^P$ & $-\frac{3}{2}$ & $-1$ & $-\frac{3}{2}$ \\
$f^S$ & $\frac{3}{2}$ & $\frac{3}{5}$ & $\frac{3}{2}$ \\
$f^A_2$ & $\frac{3}{4}$ & $\frac{1}{2}$ & $\infty$\\
$f^A_3$ & $-\infty$ & $-\frac{1}{2}$ & $-\frac{3}{4}$\\
$f^T_1$ & $0$ & $-\frac{3}{14}$ & $0$\\
$f^T_2$ & $-\frac{5}{4}$ & $-\frac{2}{7}$ & $-\frac{5}{4}$\\
$f^T_3$ & $\infty$ & $\frac{8}{21}$ & $\infty$ \\
$f^T_4$ & $\infty$ & $\frac{1}{21}$ & $-\frac{5}{12}$\\
$f^T_5$ & $-\frac{5}{12}$& $\frac{1}{21}$&$\infty$ \\
 \bottomrule 
\end{tabular}
\end{center}
\caption{Asymmetry functions evaluated at $w=0,\pm 1$. Note that none of the singularities contribute to physical helicity amplitudes. In the singly-virtual limits the overall scaling also involves factors $(1/2)^{-n}$ according to the definition of $Q^2$ in~\eqref{defQ}.}
\label{tab:asymmetry}
\end{table}

The BL scalings can be compared with the quark-model approach from~\cite{Schuler:1997yw}, whose results, translated to our notation, become
\begin{align}
\label{comparison_Schuler}
 \frac{F_{P\gamma^*\gamma^*}(q_1^2,q_2^2)}{F_{P\gamma^*\gamma^*}(0,0)}\bigg|_{\text{\cite{Schuler:1997yw}}}&=\frac{m_P^2}{m_P^2-q_1^2-q_2^2}\sim \frac{1}{Q^2},\notag\\
  \frac{\F_1^S(q_1^2,q_2^2)}{\F_1^S(0,0)}\bigg|_{\text{\cite{Schuler:1997yw}}}&=\frac{m_S^2(3m_S^2-q_1^2-q_2^2)}{3(m_S^2-q_1^2-q_2^2)^2}\sim \frac{1}{Q^2},\notag\\
 \frac{\F_2^S(q_1^2,q_2^2)}{\F_1^S(0,0)}\bigg|_{\text{\cite{Schuler:1997yw}}}&=-\frac{2m_S^4}{3(m_S^2-q_1^2-q_2^2)^2}\sim \frac{1}{Q^4},\notag\\ 
 \F_1^A(q_1^2,q_2^2)\big|_{\text{\cite{Schuler:1997yw}}}&=0,\notag\\
 \frac{\F_2^A(q_1^2,q_2^2)}{\F_2^A(0,0)}\bigg|_{\text{\cite{Schuler:1997yw}}}&=
 \frac{\F_3^A(q_1^2,q_2^2)}{\F_3^A(0,0)}\bigg|_{\text{\cite{Schuler:1997yw}}}=\bigg(\frac{m_A^2}{m_A^2-q_1^2-q_2^2}\bigg)^2\sim\frac{1}{Q^4},\notag\\
 \frac{\F_1^T(q_1^2,q_2^2)}{\F_1^T(0,0)}\bigg|_{\text{\cite{Schuler:1997yw}}}&=\bigg(\frac{m_T^2}{m_T^2-q_1^2-q_2^2}\bigg)^2\sim\frac{1}{Q^4},\notag\\
 \F_i^T(q_1^2,q_2^2)\big|_{\text{\cite{Schuler:1997yw}}}&=0,\qquad i\in\{2,3,4,5\},
\end{align}
where in all cases we have replaced the decay constants directly in terms of the TFF normalizations. In particular,  $\F_2^S(q_1^2,q_2^2)$ is indeed proportional to the normalization of $\F_1^S$ because in this framework the cross section is assumed to be proportional to the on-shell two-photon width $\Gamma_{\gamma\gamma}$ (or $\tilde\Gamma_{\gamma\gamma}$ in the case of the axial-vector mesons). Moreover, the antisymmetric part of $\F_2^A(q_1^2,q_2^2)$ is assumed to vanish, which, apart from the overall sign due to $\F_2^A(0,0)=-\F_3^A(0,0)$, makes the two non-zero axial-vector TFFs coincide. For the tensor mesons all TFFs except for $\F_1^T(q_1^2,q_2^2)$ vanish.

In all cases the non-vanishing TFFs follow the same asymptotic behavior as given in~\eqref{BL_scaling}. For the scalar TFFs, the one case in which two distinct TFFs occur, we may also check the ratio of the two, again reproducing the BL result $\F_2^S(q_1^2,q_2^2)/\F_1^S(q_1^2,q_2^2)\sim m_S^2/Q^2$ asymptotically.


\section{Comparison to data}
\label{sec:data}

With the exception of~\cite{BaBar:2018zpn} for the $\eta'$ TFF, all available data are currently restricted to singly-virtual kinematics. Moreover, while the on-shell couplings are known for a number of resonances, information on the momentum dependence is scarce, for scalar and tensor mesons the most comprehensive study comes from~\cite{Masuda:2015yoh}, addressing the $f_0(980)$ and $f_2(1270)$ resonances. 
For the axial-vector mesons, due to the Landau--Yang theorem all data are necessarily at least singly-virtual, with results available for the $f_1(1285)$~\cite{Achard:2001uu} and the $f_1'(1420)$~\cite{Achard:2007hm}. In this section, we will compare our asymptotic results to these data sets.  

\subsection{Axial-vector mesons}

The axial-vector TFFs for $A=f_1(1285), f_1'(1420)$ have been measured by the L3 collaboration in space-like $e^+e^-\to e^+e^- A$ two-photon reactions~\cite{Achard:2001uu,Achard:2007hm}, analyzed in terms of a dipole ansatz for $\F_2^A(q^2,0)$ and assuming $\F_1^A=0$
\beq
\label{L3_dipole}
\F_2^A(q^2,0)=\F_2^A(0,0)\bigg(1-\frac{q^2}{\Lambda^2}\bigg)^{-2},\qquad
\F_1^A(q^2,0)=0.
\eeq
The measured parameters are\footnote{We will assume that $\BR(K\bar K\pi)=1$ within uncertainties for the $f_1'(1420)$, given that~\cite{Tanabashi:2018oca,Armstrong:1991rg} quotes for the second-most important channel $\Gamma(\eta\pi\pi)/\Gamma(K\bar K\pi)<0.1$.}
\begin{align}
\label{L3_results}
 \tilde\Gamma_{\gamma\gamma}(f_1(1285))&=3.5(6)(5)\keV,\qquad \Lambda(f_1(1285))=1.04(6)(5)\GeV,\notag\\
 \tilde\Gamma_{\gamma\gamma}(f_1'(1420))\BR(K\bar K\pi)&=3.2(6)(7)\keV,\qquad \Lambda(f_1'(1420))=0.926(72)(31)\GeV.
\end{align}
Further, the analysis is based on the cross section
\begin{align}
 \sigma_{\gamma^*\gamma\to A}&=2\pi^2\alpha^2\frac{m_A\Gamma_A}{(s-m_A^2)^2+m_A^2\Gamma_A^2}\bigg(1-\frac{q^2}{m_A^2}\bigg)\notag\\
 &\qquad\times\Bigg[\bigg|\bigg(1-\frac{q^2}{m_A^2}\bigg)\F_1^A(q^2,0)-\frac{q^2}{m_A^2}\F_2^A(q^2,0)\bigg|^2-\frac{2q^2}{m_A^2}\big|\F_2^A(q^2,0)\big|^2\Bigg]\notag\\
 &\overset{\F_1^A\to0}{=}24\pi\frac{\Gamma_A\tilde \Gamma_{\gamma\gamma}}{(s-m_A^2)^2+m_A^2\Gamma_A^2}\bigg(1-\frac{q^2}{m_A^2}\bigg)\frac{-q^2}{m_A^2}\bigg(2-\frac{q^2}{m_A^2}\bigg)\bigg|\frac{\F_2^A(q^2,0)}{\F_2^A(0,0)}\bigg|^2,
 \label{sigmaggA}
\end{align}
where the simplification for $\F_1^A=0$ reproduces the expression in~\cite{Achard:2001uu}. Unfortunately, the original data for $\sigma_{\gamma^*\gamma\to A}$ cannot be extracted from~\cite{Achard:2001uu,Achard:2007hm}, accordingly, we will compare to the band for $\F_2^A$ given by the dipole ansatz~\eqref{L3_dipole}.
Defining an effective decay constant by
\beq
F_A^\text{eff}=4\sum_a C_a F_A^a,
\eeq
we have the asymptotic limits
\beq
\label{axial_vector_asym}
\F_2^A(q^2,q^2)=\frac{F_A^\text{eff}m_A^3}{2q^4}+\Order\big(q^{-6}\big),\qquad \F_2^A(q^2,0)=\frac{3F_A^\text{eff}m_A^3}{q^4}+\Order\big(q^{-6}\big),
\eeq
and
\beq
\label{axial_vector_mass}
\F_2^A(q^2,0)=\frac{3F_A^\text{eff}m_A^3}{q^4}\times \frac{2}{x^2}\bigg(\frac{x}{1-x}+\log(1-x)\bigg),\qquad x=\frac{m_A^2}{q^2},
\eeq
when keeping the axial-vector mass in~\eqref{result_final}.

Since additional phenomenological input that could constrain $F_A^\text{eff}$ is scarce, we will now consider these decay constants as have been estimated using light-cone sum rules (LCSRs)~\cite{Yang:2007zt}. In particular, results are provided for the $a=0,3,8$ components, but to extract $F_A^\text{eff}$ for the physical mesons, mixing effects need to be taken into account. We introduce the mixing angle $\theta_A$ via
\beq
\begin{pmatrix}
 f_1\\ f_1'
\end{pmatrix}
=\begin{pmatrix}
 \cos\theta_A & \sin\theta_A\\
 -\sin\theta_A & \cos\theta_A
\end{pmatrix}
\begin{pmatrix}
 f^0\\ f^8
\end{pmatrix},
\eeq
in terms of which
\beq
\label{mixing_axial_vector}
\frac{\tilde \Gamma_{\gamma\gamma}(f_1)}{\tilde \Gamma_{\gamma\gamma}(f_1')}=\frac{m_{f_1}}{m_{f_1'}}\cot^2(\theta_A-\theta_0),\qquad \theta_0=\arcsin\frac{1}{3}.
\eeq
$\theta_0$ is the mixing angle that leads to a vanishing two-photon coupling of $f_1'$. Octet/singlet mixing is reproduced for $\theta_A=\pi/2$, ideal mixing for $\theta_A=\arctan 1/\sqrt{2}=35.3\degree$, and the L3 results~\eqref{L3_results} imply 
$\theta_A=62(5)\degree$~\cite{Achard:2007hm}.
Further, we can use $SU(3)$ symmetry to extract an empirical width for the $a_1(1260)$
\beq
\tilde \Gamma_{\gamma\gamma}(a_1)=\frac{\tilde \Gamma_{\gamma\gamma}(f_1)}{3\cos^2(\theta_A-\theta_0)}\frac{m_{a_1}}{m_{f_1}}
=m_{a_1}\frac{m_{f_1} \tilde \Gamma_{\gamma\gamma}(f_1')+m_{f_1'} \tilde \Gamma_{\gamma\gamma}(f_1)}{3m_{f_1} m_{f_1'}}=2.0(7)\keV,
\eeq
where we added in quadrature the uncertainties from $\tilde \Gamma_{\gamma\gamma}(f_1)$, $\tilde \Gamma_{\gamma\gamma}(f_1')$, $m_{a_1}$, as well as a generic $30\%$ $SU(3)$ uncertainty.

Denoting the decay constants and masses in Cartesian basis by $F_A^a$ and $m^a_A$, we obtain for the decay constants parameterizing the $q=u,d,s$ currents
\begin{align}
 F_{f_1}^u&=F_{f_1}^d=F^0_A\sqrt{\frac{2}{3}}\frac{m^0_A}{m_{f_1}}\cos\theta_A + \frac{F^8_A}{\sqrt{3}}\frac{m^8_A}{m_{f_1}}\sin\theta_A,
 & F_{f_1}^s&=F^0_A\sqrt{\frac{2}{3}}\frac{m^0_A}{m_{f_1}}\cos\theta_A - \frac{2F^8_A}{\sqrt{3}}\frac{m^8_A}{m_{f_1}}\sin\theta_A,\notag\\
 F_{f_1'}^u&=F_{f_1'}^d=-F^0_A\sqrt{\frac{2}{3}}\frac{m^0_A}{m_{f_1'}}\sin\theta_A + \frac{F^8_A}{\sqrt{3}}\frac{m^8_A}{m_{f_1'}}\cos\theta_A,
 & F_{f_1'}^s&=-F^0_A\sqrt{\frac{2}{3}}\frac{m^0_A}{m_{f_1'}}\sin\theta_A - \frac{2F^8_A}{\sqrt{3}}\frac{m^8_A}{m_{f_1'}}\cos\theta_A,\notag\\
 F_{a_1}^u&=-F_{a_1}^d=F^3_A, & F_{a_1}^s&=0,
\end{align}
where we have further assumed isospin symmetry and allowed for the physical masses of the $f_1$ and $f_1'$ to differ from the singlet and octet ones. 
The relations for $f_1$ and $f_1'$ differ by a factor of $\sqrt{2}$ from~\cite{Yang:2007zt}, which leads us to the identification
\beq
 \sqrt{2} F_A^0=245(13)\MeV, \qquad \sqrt{2} F_A^8=239(13)\MeV,\qquad \sqrt{2} F_A^3=238(10)\MeV.
\eeq
Together with the masses $m_A^0=1.28(6)\GeV$ and $m_A^8=1.29(5)\GeV$, this leads to
\begin{align}
\label{LCSR_result}
F_{f_1}^\text{eff}&=2F^0_A\bigg(\frac{2}{3}\bigg)^{3/2}\frac{m^0_A}{m_{f_1}}\cos\theta_A
+\frac{2F^8_A}{3\sqrt{3}}\frac{m^8_A}{m_{f_1}}\sin\theta_A=146(7)(12)\MeV,\notag\\
F_{f_1'}^\text{eff}&=-2F^0_A\bigg(\frac{2}{3}\bigg)^{3/2}\frac{m^0_A}{m_{f_1'}}\sin\theta_A
+\frac{2F^8_A}{3\sqrt{3}}\frac{m^8_A}{m_{f_1'}}\cos\theta_A=-122(11)(11)\MeV,\notag\\
 F_{a_1}^\text{eff}&=\frac{2}{3}F_A^3=112(5)\MeV,
\end{align}
where the first uncertainty is propagated from the LCSRs, while the second refers to the uncertainty in the mixing angle. 
We note that in all cases the effective decay constants $F_A^\text{eff}=\F_2^A(0,0) m_A/2$ suggested by~\cite{Schuler:1997yw}, when matching in the doubly-virtual direction~\eqref{comparison_Schuler}, exceed the LCSR estimates by about a factor $2$, indicating that the quark model overestimates the asymptotic coefficients.\footnote{When matching in the singly-virtual direction the mismatch would reduce because instead of the relative factor $6$ as in~\eqref{axial_vector_asym} the quark model only has a factor $4$. However, in both cases the doubly-virtual prediction is expected to be more reliable. For this comparison, we adjust the normalization of the quark model to the L3 data.} Finally, extrapolating the dipole fit~\eqref{L3_dipole} would imply an even lower coefficient
\beq
F_{f_1}^\text{eff}=82(26)\MeV,\qquad F_{f_1'}^\text{eff}=-34(12)\MeV,
\eeq
but in both cases there is only a single bin above $1\GeV^2$, rendering conclusions about the asymptotics highly uncertain. 

Beyond LCSRs, the effective decay constant $F_{a_1}^\text{eff}$ can, in principle, be extracted from $\tau\to 3\pi \nu_\tau$ decays. Such extractions typically lead to $F_{a_1}^\text{eff}=(95\ldots 100)\MeV$~\cite{Isgur:1988vm,Dumm:2009va}, in reasonable agreement with the LCSR value in~\eqref{LCSR_result}, but the systematic uncertainties due to the $a_1$ spectral shape are substantial. In contrast, as isospin singlets the neutral $f_1$, $f_1'$ cannot be produced in $\tau$ decays. Further, there is an early lattice-QCD calculation that quotes $F_{a_1}^\text{eff}=113(13)\MeV$~\cite{Wingate:1995hy}, while more recent calculations of the $a_1$ have concentrated on mass and width~\cite{Lang:2014tia,Murakami:2018spb}. Especially for $f_1$ and $f_1'$, additional input would be highly welcome, as it would remove the main uncertainty in the asymptotic BL relations.

\begin{figure}[t]
	\centering
	\includegraphics[width=0.49\linewidth]{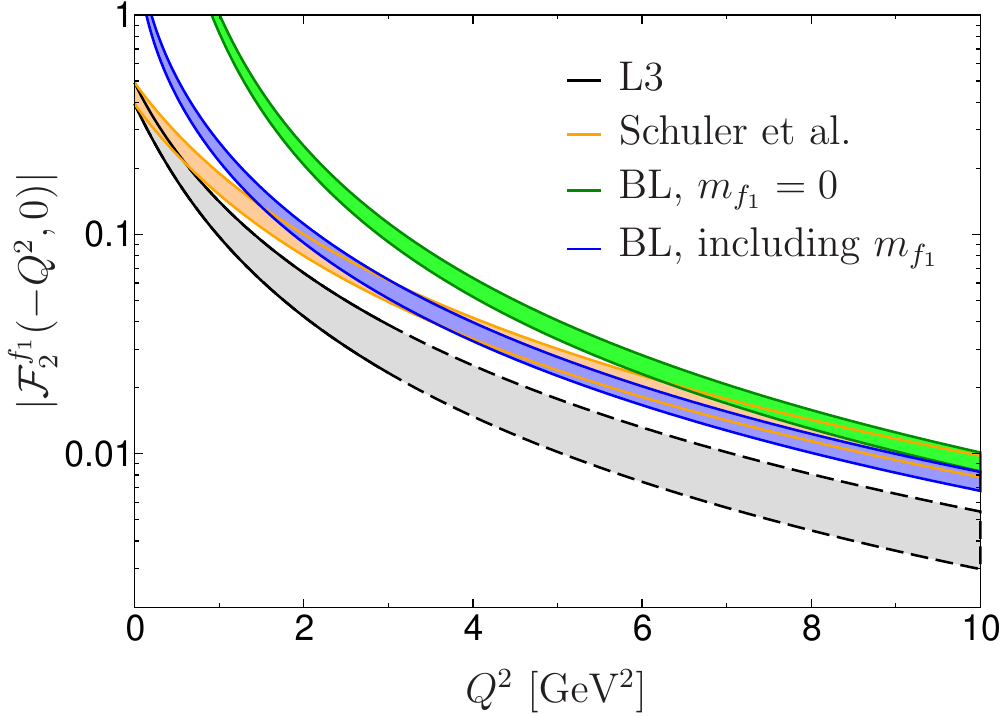}
	\includegraphics[width=0.49\linewidth]{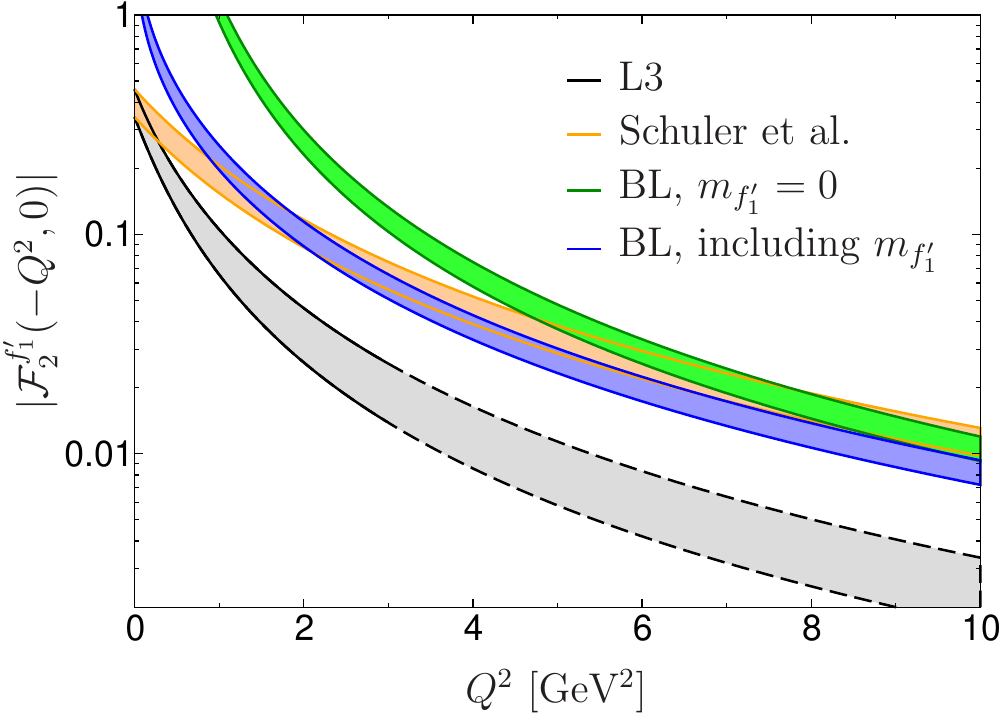}
	\caption{Axial-vector TFF $\F_2^A$ for $f_1(1285)$ (left) and $f_1'(1420)$ (right). In each case, the gray band refers to the dipole fit~\eqref{L3_dipole} with parameters~\eqref{L3_results}, the orange band to the quark model from~\cite{Schuler:1997yw}, see~\eqref{comparison_Schuler} (with normalization adjusted to the L3 data), the green band to the asymptotic BL result~\eqref{axial_vector_asym}, and the blue band to the variant including the axial-vector mass~\eqref{axial_vector_mass}. The uncertainties are propagated from~\eqref{L3_results} and~\eqref{LCSR_result}, respectively. The L3 dipole fit is indicated by dashed lines above $3\GeV^2$ (close to the center of the last bin), to emphasize the fact that only a single bin probes the region above $1\GeV^2$.}      
	\label{fig:L3}
\end{figure}

The comparison to the L3 dipole fit is shown in Fig.~\ref{fig:L3}. In both cases the quark-model result decreases more slowly than the BL bands, but especially for the $f_1'$ both quark model and BL lie significantly above the extrapolated L3 fit. However, the fit is dominated by the bins below $1\GeV^2$, while mass corrections are important well beyond, as indicated by the comparison of the two BL bands. In addition, while $\F_1^A$ is suppressed both for small virtualities (its symmetry properties require $\F_1^A(-Q^2,0)\sim Q^2$) and for large virtualities ($\F_1^A(-Q^2,0)\sim 1/Q^6$ according to~\eqref{result_final}), there may still be a significant contribution for intermediate virtualities, which by means of the relative signs in~\eqref{sigmaggA} could indeed effectively suppress the results for $\F_2^A$ extracted under the assumption~\eqref{L3_dipole}.

\subsection{Scalar and tensor mesons}

The singly-virtual TFFs for scalar and tensor mesons have been studied using light-cone methods in~\cite{Kroll:2016mbt} and~\cite{Braun:2016tsk}, respectively, including terms beyond the asymptotic results we considered here. We refer to these works regarding the potential impact of these subleading contributions, but show here how the leading terms compare to phenomenology. 

\begin{figure}[t]
	\centering
	\includegraphics[width=0.49\linewidth]{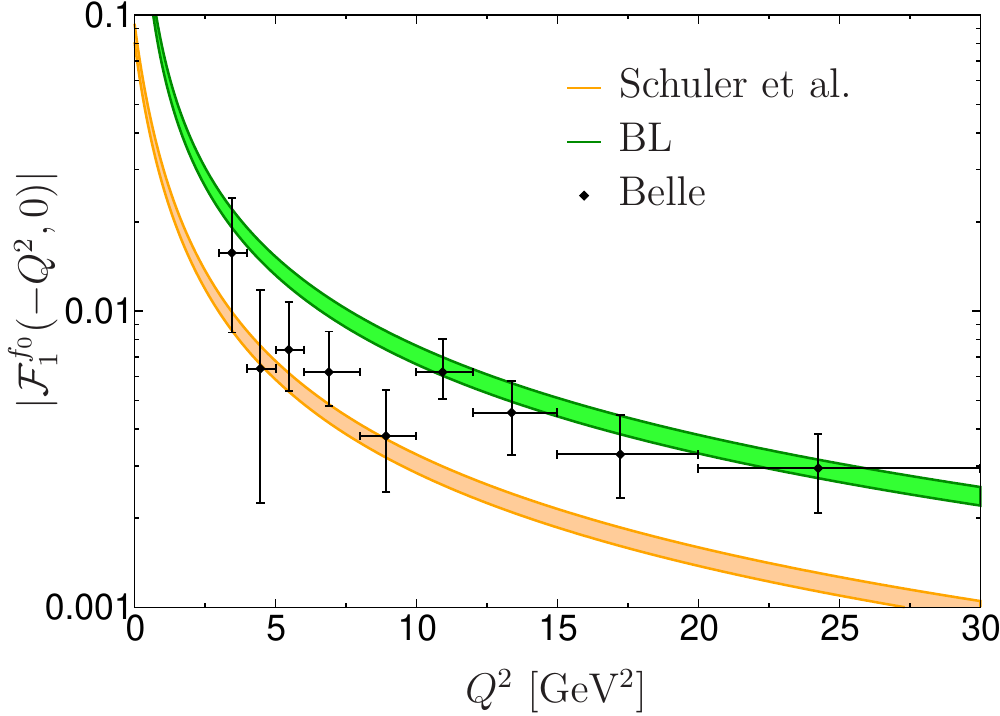}
	\caption{Scalar TFF $\F_1^S$ for the $f_0(980)$, in comparison to the Belle data~\cite{Masuda:2015yoh}. 
	The orange band refers to the quark model from~\cite{Schuler:1997yw}, see~\eqref{comparison_Schuler}, and the green band 
	to the asymptotic BL result~\eqref{scalar_asym}, with effective decay constant determined by matching to~\cite{Schuler:1997yw} in the doubly-virtual direction. In both cases, the uncertainties are propagated from $\Gamma_{\gamma\gamma}$~\cite{Tanabashi:2018oca}.}      
	\label{fig:Bellescalar}
\end{figure}

For the scalar mesons in the singly-virtual limit only the helicity amplitude $H_{++}$ is relevant, and therein only the contribution from $\F_1^S$. Accordingly, the results for the $f_0(980)$ in~\cite{Masuda:2015yoh} can be interpreted as $\F_1^S(-Q^2,0)/\F_1^S(0,0)$, where for the normalization a two-photon width $\Gamma_{\gamma\gamma}=0.29^{+0.07}_{-0.06}\keV$ and $m_{f^0}=0.98\GeV$ were assumed. With this input, we can reconstruct the data points for $\F_1^S(-Q^2,0)$. For a definite comparison to the BL result one would need independent input for the effective decay constant
\beq
F_S^\text{eff}=4\sum_a C_a \bar F_S^a(\mu) B_1(\mu).
\eeq
Absent such information, we can again match to~\cite{Schuler:1997yw} in the doubly-virtual direction, which gives
\beq
F_S^\text{eff}=\frac{5}{18}\F_1^S(0,0)m_S,
\eeq
and thus $F_{f_0}^\text{eff}=24(2)\MeV$ (using current PDG numbers $\Gamma_{\gamma\gamma}=0.31^{+0.05}_{-0.04}\keV$, $m_{f^0}=0.99(2)\GeV$~\cite{Tanabashi:2018oca}), while the result for the matching in the singly-virtual direction would be lower by a factor $5/2$. In Fig.~\ref{fig:Bellescalar} we show the comparison to the resulting
\beq
\label{scalar_asym}
\F_1^S(-Q^2,0)=\frac{3F_S^\text{eff}m_S}{Q^2},
\eeq
which asymptotically indeed indicates better agreement with the data for the doubly-virtual matching. 

\begin{figure}[t]
	\centering
	\includegraphics[width=0.49\linewidth]{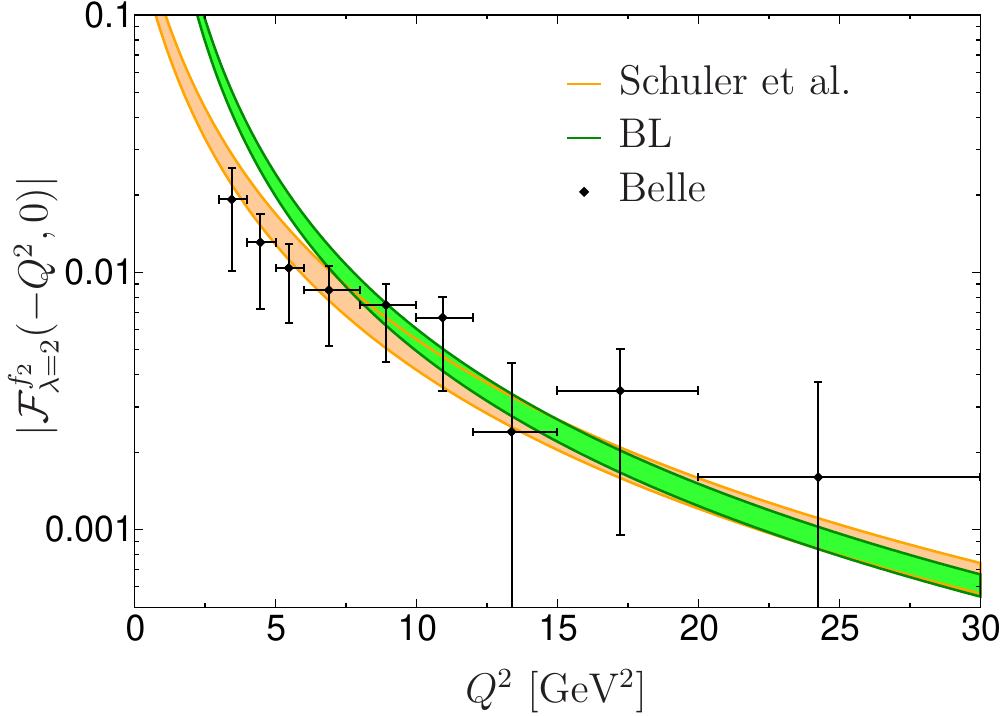}
	\includegraphics[width=0.49\linewidth]{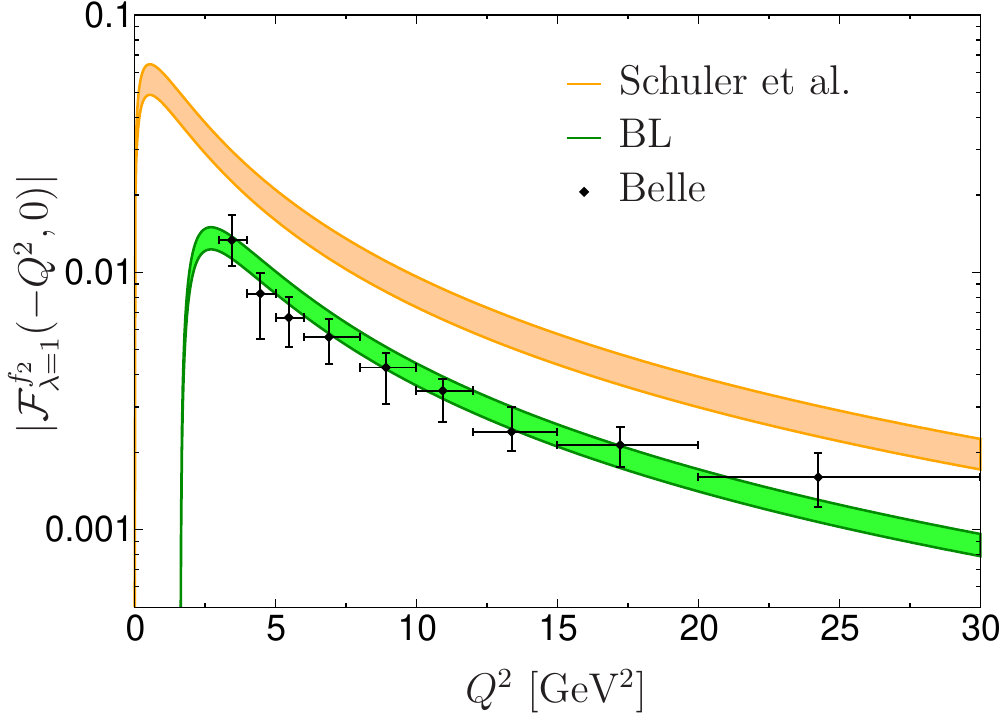}\\[0.2cm]
	\includegraphics[width=0.49\linewidth]{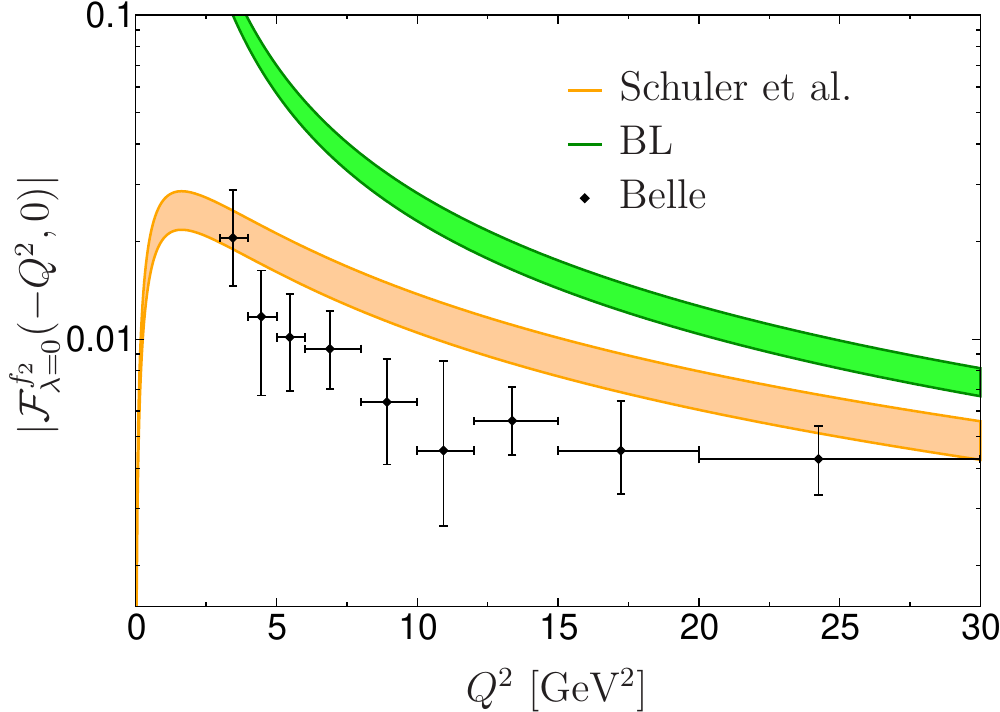}
	\caption{Tensor TFFs for helicities $\lambda=2,1,0$ for the $f_2(1270)$, in comparison to the Belle data~\cite{Masuda:2015yoh}. 
	The orange band refers to the quark model from~\cite{Schuler:1997yw}, see~\eqref{comparison_Schuler}, and the green band 
	to the asymptotic BL result~\eqref{tensor_asym}, with effective decay constant from~\eqref{FT_LCSR}. For the quark-model normalization uncertainties are propagated from $\Gamma_{\gamma\gamma}$~\cite{Tanabashi:2018oca}, assuming that this also covers the contribution from $\F_2^T(0,0)$ in~\eqref{onshell_tensor}.}      
	\label{fig:Belletensor}
\end{figure}

For the comparison of the tensor TFFs, we first need to map conventions. The results in~\cite{Masuda:2015yoh} are presented in helicity basis, and according to~\eqref{hel_tensor} this probes the linear combinations
\begin{align}
\label{FThel}
 \F^T_{\lambda=0}(-Q^2,0)&=\frac{Q^2}{\sqrt{6}m_T^2}\F_1^T(-Q^2,0)-\frac{(m_T^2+Q^2)^2}{2\sqrt{6} m_T^4} \F_2^T(-Q^2,0)+\frac{Q^2}{\sqrt{6}m_T^2}\F_5^T(-Q^2,0),\notag\\
 \F^T_{\lambda=1}(-Q^2,0)&=\frac{\sqrt{Q^2}}{\sqrt{2}m_T}\F_1^T(-Q^2,0)+\frac{\sqrt{Q^2}(m_T^2-Q^2)}{2\sqrt{2} m_T^3} \F_5^T(-Q^2,0),\notag\\
 \F^T_{\lambda=2}(-Q^2,0)&=-\F_1^T(-Q^2,0)+\frac{Q^2}{m_T^2}\F_5^T(-Q^2,0).
\end{align}
Moreover, the normalization of the results accounts for the small contribution from $\F_2(0,0)$ to $\Gamma_{\gamma\gamma}$, see~\eqref{onshell_tensor}, so that the full results are restored by multiplication with $\sqrt{5\Gamma_{\gamma\gamma}/(\pi\alpha^2 m_T)}$ with $\Gamma_{\gamma\gamma}=3.0(4)\keV$.
Finally, the data only provide information on the absolute values, but not the relative signs, so that an explicit inversion for the $\F_i^T$ requires assumptions on these relative phases. For this reason, we will work directly with the helicity combinations~\eqref{FThel}, in terms of which the BL constraints become
\begin{align}
 \F^T_{\lambda=0}(-Q^2,0)&=-\frac{5F_T^\text{eff}m_T}{3\sqrt{6}Q^6}\big(3Q^4+4m_T^2Q^2+3m_T^4\big),\notag\\
 \F^T_{\lambda=1}(-Q^2,0)&=-\frac{5\sqrt{2}F_T^\text{eff}m_T^2(Q^2-m_T^2)}{6Q^5},\notag\\
 \F^T_{\lambda=2}(-Q^2,0)&=\frac{10F_T^\text{eff}m_T^3}{3Q^4},
 \label{tensor_asym}
\end{align}
with effective decay constant
\beq
F_T^\text{eff}=4\sum_a C_a F_T^a.
\eeq
The non-strangeness components have been estimated from LCSRs in~\cite{Aliev:1981ju,Cheng:2010hn,Braun:2016tsk}, which provides by far the dominant contribution given that the $f_2(1270)$--$f_2'(1525)$ system is close to ideal mixing.\footnote{Using $\Gamma_{\gamma\gamma}(f_2)=2.6(5)\keV$, $\Gamma_{\gamma\gamma}(f_2')=0.081(9)\keV$~\cite{Tanabashi:2018oca}, the analog of~\eqref{mixing_axial_vector} gives $\theta_T=29(1)\degree$, indeed very close to $\arctan 1/\sqrt{2}=35.3\degree$.} Numerically, we will use~\cite{Braun:2016tsk}
\beq
\label{FT_LCSR}
F_T^\text{eff}=\frac{5}{9}\sqrt{2}F_T^q=79(8)\MeV.
\eeq
In this case, we do not attempt to match to the quark model, given that the structure of the tensor amplitudes is fundamentally different: in~\cite{Schuler:1997yw}, all $\F_i^T$ except for $\F_1^T$ vanish, while in the BL case it is precisely $\F_1^T$ that vanishes in the singly-virtual limit. Even for doubly-virtual kinematics the coefficient is very small, see Table~\ref{tab:asymmetry}, so that the matching to~\eqref{comparison_Schuler} would lead to $F_T^\text{eff}$ almost a factor $5$ above the LCSR estimate. The comparison to the data is shown in Fig.~\ref{fig:Belletensor}. It is quite remarkable that the helicity-$2$ form factor is well described in either formalism, given that the contributions originate from completely different Lorentz structures. That is, in the quark model the vanishing TFFs $\F_{2,5}^T$ are compensated by $\F_1^T$. For the helicity-$1$ form factor we observe excellent agreement between data and the BL result, while for the helicity-$0$ projection the asymptotic behavior appears to set in rather late. The agreement in the helicity-$0$ TFF seems to improve when including subleading corrections~\cite{Braun:2016tsk}, but the uncertainties associated with the additional matrix elements are substantial.

\section{Summary and outlook}
\label{sec:summary}

In this paper we studied the asymptotic behavior of meson TFFs as motivated by resonance contributions to HLbL scattering in $(g-2)_\mu$. 
To this end, we first applied the BTT procedure to the two-photon matrix elements of pseudoscalar, scalar, axial-vector, and tensor mesons to obtain a gauge-invariant Lorentz decomposition that is demonstrably free of kinematic singularities. Using light-cone distribution amplitudes from the literature, we then derived the leading asymptotic behavior for the TFFs that emerge in the BTT decomposition and compared the results to quark-model expectations. For the axial-vector mesons we compared to the available phenomenological information on the singly-virtual process from L3, which, however, does not suffice to conclusively challenge the prediction for the asymptotic coefficient obtained when combining the BL limit with LCSR estimates of the decay constants. In addition, we compared the asymptotic results for scalar and tensor mesons to singly-virtual data from Belle. 
In all cases, the main uncertainty in the asymptotic coefficient arises from limited knowledge of the decay constants, which in principle could be calculated in lattice QCD.     

The results presented here provide valuable constraints on the TFFs required to estimate the contribution from multi-hadron channels to HLbL scattering in terms of narrow resonances. In close analogy to the pseudoscalar poles, information about the asymptotic behavior is necessary to assess the impact of the high-energy tails in the $(g-2)_\mu$ integral. Here, we derived the corresponding limits for scalar, axial-vector, and tensor mesons, as well as suitable Lorentz decompositions that avoid introducing kinematic singularities, contrary to decompositions into definite helicity components. 
In particular, we expect that our results will facilitate improved estimates for the contribution from intermediate energies around $1$--$2\GeV$ to HLbL scattering, to help further elucidate the critical interplay of exclusive hadronic channels, resonance contributions, and short-distance constraints.

\section*{Acknowledgements}
\addcontentsline{toc}{section}{Acknowledgements}

We thank A.~Khodjamirian and B.~Kubis for useful discussions, S.~Mele for communication regarding~\cite{Achard:2001uu,Achard:2007hm}, 
and G.~Colangelo, F.~Hagelstein, and M.~Procura for comments on the manuscript. 
Financial support by the Swiss National Science Foundation (Project No.~PCEFP2\_181117)
and the DOE (Grants No.~DE-FG02-00ER41132 and No.~DE-SC0009919) is gratefully acknowledged.
P.\,S.\ thanks the INT at the University of Washington for its hospitality and the DOE for partial support during an early stage of this work.

\renewcommand\bibname{References}
\renewcommand{\bibfont}{\raggedright}
\bibliographystyle{utphysmod}
\phantomsection
\addcontentsline{toc}{section}{References}
\bibliography{Literature}

\end{document}